 \def\draftversion{false}
\RequirePackage{ifthen}
\ifthenelse{\equal{\draftversion}{true}}{
  \documentclass[aps,pra,10pt,galley,amsmath,amssymb,nofootinbib,
     superscriptaddress]{revtex4}
}{
  \documentclass[aps,pra,10pt,twocolumn,amsmath,amssymb,
     superscriptaddress,longbibliography,nofootinbib]{revtex4-1}
}

\usepackage{graphicx}% Include figure files
\usepackage[usenames,dvipsnames]{color} % colors
\usepackage{bm} % bold math
\usepackage{soul} % \st    for strike-out
\usepackage{mathtools}
\usepackage{graphicx}% Include figure files
\usepackage[update,prepend]{epstopdf}
\usepackage{dcolumn} % allow decimal alignment in tables
\usepackage{comment}
\usepackage{gensymb}
\usepackage[linktocpage=true]{hyperref}
\usepackage{hyperref}
\usepackage{comment}
\usepackage{notes2bib}
\usepackage{multirow}
\hypersetup{
  pdfnewwindow=true, colorlinks=true,
  linkcolor=blue, anchorcolor=blue,
  citecolor=blue, filecolor=blue,
  menucolor=blue, urlcolor=blue}

\def\mscolor{Orange}
\def\srcolor{Red}
\def\jbcolor{OliveGreen}
\def\cdcolor{blue}
\def\dvcolor{Magenta}
\newcounter{comm} % counter for margin comments
\newcommand{\margin}[3]{\stepcounter{comm}
  \marginpar{\small \textcolor{#1}{#2[\arabic{comm}]: #3}}
  \textcolor{#1}{[\arabic{comm}]}}
\newcommand{\srm}[1]{\margin{\srcolor}{SR}{#1}}
\newcommand{\jbm}[1]{\margin{\jbcolor}{JB}{#1}}
\newcommand{\cdm}[1]{\margin{\cdcolor}{CD}{#1}}
\newcommand{\msm}[1]{\margin{\mscolor}{MS}{#1}}
\newcommand{\dvm}[1]{\margin{\dvcolor}{DV}{#1}}

\ifthenelse{\equal{\draftversion}{true}}{
  \marginparwidth 2.7in
  \marginparsep 0.5in
  \def\mlab#1{\marginpar{\small\bf #1}}
  
  \newcommand{\seclab}[1]{\label{sec:#1}{\Red{\small\;\;[sec:~#1]}}}
  \newcommand{\aplab}[1]{\label{ap:#1}{\Red{\small\;\;[ap:~#1]}}}
  \newcommand{\eqlab}[1]{\Red{\hbox{\small\;\;[#1]}}\label{eq:#1}}
  \newcommand{\figlab}[1]{\Red{\hbox{\small\;\;[fig:~#1]}}\label{fig:#1}}
}{
  \renewcommand{\srm}[1]{}
  \renewcommand{\jbm}[1]{}
  \renewcommand{\cdm}[1]{}
  \renewcommand{\msm}[1]{}
  \renewcommand{\dvm}[1]{}
  \def\mlab#1{}
  
  \newcommand{\eqlab}[1]{\label{eq:#1}}
  \newcommand{\seclab}[1]{\label{sec:#1}}
  \newcommand{\aplab}[1]{\label{ap:#1}}
  \newcommand{\figlab}[1]{\label{fig:#1}}
}

\newcommand{\beq}{\begin{equation}}
\newcommand{\eeq}{\end{equation}}
\newcommand{\bea}{\begin{eqnarray}}
\newcommand{\eea}{\end{eqnarray}}
\newcommand{\nn}{\nonumber\\}
\newcommand{\eq}[1]{Eq.~(\ref{eq:#1})}
\newcommand{\Eq}[1]{Equation~(\ref{eq:#1})}
\newcommand{\eqs}[2]{Eqs.~(\ref{eq:#1}) and (\ref{eq:#2})}

\newcommand{\eqr}[2]{Eqs.~(\ref{eq:#1}-\ref{eq:#2})}

\newcommand{\fref}[1]{Fig.~\ref{fig:#1}}
\newcommand{\Fref}[1]{Figure~\ref{fig:#1}}

\newcommand{\sref}[1]{Sec.~\ref{sec:#1}}
\newcommand{\srefs}[2]{Secs.~\ref{sec:#1} and \ref{sec:#2}}
\newcommand{\Sref}[1]{Section~\ref{sec:#1}}
\newcommand{\ket}[1]{\vert#1\rangle}
\newcommand{\bra}[1]{\langle#1\vert}
\newcommand{\ip}[2]{\langle#1\vert#2\rangle}
\newcommand{\me}[3]{\langle#1\vert#2\vert#3\rangle}
\newcommand{\ev}[1]{\langle#1\rangle}
\def\Im{\mathrm{Im}}
\def\Tr{\mathrm{Tr}}
\def\z2{$\mathbb{Z}_2$}

\def\phm{\phantom{-}}

\newcommand{\abs}[1]{\vert#1\vert}

\def\q{\mathbf q}
\def\rp{\mathrm{p}}

\def\rN{\mathrm{N}}
\def\re{\mathrm{e}}
\def\kpp{K^{\mathrm{(pp)}}}
\def\ksp{K^{\mathrm{(sp)}}}
\def\kps{K^{\mathrm{(ps)}}}
\def\kss{K^{\mathrm{(ss)}}}

\def\gpp{G^{\mathrm{(pp)}}}
\def\gsp{G^{\mathrm{(sp)}}}
\def\gps{G^{\mathrm{(ps)}}}
\def\gss{G^{\mathrm{(ss)}}}

\def\opp{\Omega^{\mathrm{(pp)}}}
\def\osp{\Omega^{\mathrm{(sp)}}}

\def\oss{\Omega^{\mathrm{(ss)}}}

\def\Mp{M^{\mathrm{(pp)}}}

\def\ve{\varepsilon\phantom{^*}}

\begin{document}

%===========================%
% TITLE PAGE                %
%===========================%

\title{Adiabatic dynamics of coupled spins and phonons in magnetic insulators}

\author{Shang Ren}
\affiliation{Department of Physics and Astronomy, Rutgers University, Piscataway, New Jersey 08854, USA}
\affiliation{Center for Computational Quantum Physics, Flatiron Institute, 162 5th Avenue, New York, New York 10010,
USA}

\author{John Bonini}
\affiliation{Center for Computational Quantum Physics, Flatiron Institute, 162 5th Avenue, New York, New York 10010,
USA}

\author{Massimiliano Stengel}
\affiliation{Institut de Ci\`{e}ncia de Materials de Barcelona (ICMAB-CSIC), Campus UAB, 08193 Bellaterra, Spain}
\affiliation{ICREA-Instituci\'{o} Catalana de Recerca i Estudis Avan\c{c}ats, 08010 Barcelona, Spain}

\author{Cyrus E. Dreyer}
\affiliation{Department of Physics and Astronomy, Stony Brook University, Stony Brook, New York, 11794-3800,
  USA}
\affiliation{Center for Computational Quantum Physics, Flatiron Institute, 162 5th Avenue, New York, New York 10010,
USA}

\author{David Vanderbilt}
\affiliation{Department of Physics and Astronomy, Rutgers University, Piscataway, New Jersey 08854, USA}

% See definition above
%\mydate

\begin{abstract}

In conventional \textit{ab initio} methodologies, phonons are calculated by solving equations of motion involving static interatomic force constants and atomic masses. The Born-Oppenheimer approximation, where all electronic degrees of freedom are assumed to adiabatically follow the nuclear dynamics, is also adopted. This approach does not fully account for the effects of broken time-reversal symmetry in systems with magnetic order. Recent attempts to rectify this involve the inclusion of the velocity dependence of the interatomic forces in the equations of motion, which accounts for time-reversal symmetry breaking, and can result in chiral phonon modes with non-zero angular momentum even at the zone center. However, since the energy ranges of phonons and magnons typically overlap, the spins cannot be treated as adiabatically following the lattice degrees of freedom.  Instead, phonon and spins must be treated on a similar footing. Focusing on zone-center modes, we propose a method involving Hessian matrices and Berry curvature tensors in terms of both phonon and spin degrees of freedom, and describe a first-principles methodology for calculating these. We then solve Lagrange's equations of motion to determine the energies and characters of the mixed excitations, allowing us to quantify, for example, the energy splittings between chiral pairs of phonons in some cases, and the degree of magnetically induced mixing between infrared and Raman modes in others. The approach is general, and can be applied to determine the adiabatic dynamics of any mixed set of slow variables. 

\end{abstract}

\maketitle

%===========================%
% MAIN TEXT                 %
%===========================%

%=================================================
\section{Introduction}
\seclab{intro}
%=================================================

An outstanding challenge of first-principles materials theory is the development of a systematic treatment of the coupled dynamics of phonons and magnons in magnetic materials. The calculation of phonon dispersions has long been a standard feature of modern density-functional theory (DFT) codes, based either on finite-difference or linear-response calculations of the dynamical matrix~\cite{gonze1997,baroni2001,phonopy-phono3py-JPCM,giustino2017,Monserrat2018}. On the other hand, magnon dispersions are most often computed in the context of discrete spin models, sometimes using parameters derived from DFT~\cite{Liechtenstein1987,Halilov1998,Ke2021,Durhuus2023}, although treatments based on time-dependent DFT (TDDFT)~\cite{Savrasov1998,Lounis2011,Rousseau2012,Wysocki2017,Cao2018,Tancogne2020,Skovhus2021,Gorni2022} and many-body perturbation theory~\cite{Aryasetiawan1999,Karlsson200,Kotani2008,Sasioglu2010,Muller2019} have also appeared. However, the consistent treatment of phonon and magnon dynamics on a similar footing, and the coupling between them, remains daunting \cite{delugas2023}.

Phonons play a crucial role in determining various thermodynamic and electronic properties of materials, including heat capacity, heat transport, electronic conductivity, and superconductivity. Conventionally, phonons are treated within the Born-Oppenheimer approximation~\cite{born1954}, i.e., assuming that the electronic degrees of freedom (DOF) can adiabatically follow the motion of the atoms.
In these calculations, the potential energy in the phonon Hamiltonian is computed as a function of atomic displacements. The usual harmonic approximation involves keeping only the leading quadratic dependence of the energy on displacement, encoded in the interatomic force constant (IFC) matrix, i.e., the Hessian matrix of the energy. Anharmonic treatments go further by taking account of higher-order tensors which describe third and higher derivatives with respect to displacements. These harmonic and anharmonic tensors are all invariant under time-reversal symmetry (TRS), so that at this level of description the phonons are assumed to preserve TRS and to possess the symmetries of the nonmagnetic group.

However, this assumption is incorrect for materials with magnetic order or in the presence of an external magnetic field. Recent efforts have been made to address this issue by incorporating the nuclear Berry potential into the effective Hamiltonian~\cite{mead1979,zhang2014,chen2019,coh2021,komiyama2021,bistoni2021,saparov2022,wang2022,bonini2023}, which arises naturally when the electronic DOF are integrated out under the Born-Oppenheimer approximation, as first pointed out by Mead and Truhlar~\cite{mead1979}.
The nuclear Berry potential introduces a velocity-dependent force into the equations of motion (EOM), which is determined by the nuclear Berry curvature multiplied by the nuclear velocity.

Theoretical studies have made several predictions regarding the impact of the nuclear Berry curvature on zone-center chiral phonons~\cite{zhang2014,chen2019,coh2021,komiyama2021, saparov2022,wang2022,bonini2023}, and their contribution to the thermal Hall effect~\cite{zhang2010,qin2012,zhang2016,saito2019}. Experimental evidence has also emerged supporting the existence of zone-center chiral phonons~\cite{du2019,yin2021} and the phonon thermal Hall effect~\cite{grissonnanche2019,zhang2021}. 
We define chiral phonons as those that respect rotational symmetry but possess complex eigenvalues under the rotation operator. 
Chiral phonons have attracted attention due to their unique properties, such as their selective excitation by circular-polarized light with different helicities~\cite{chen2015,du2019,chen2019np,yin2021}, their coupling to electronic states with distinct chiralities~\cite{zhu2018} following the selection rule proposed in Ref.~\cite{zhang2015}, and their ability to demonstrate Floquet behavior when driven by lasers~\cite{hubener2018}. Additionally, chiral phonons can possess phonon magnetic moments~\cite{ren2021,juraschek2022,zhang2023}.

To date, much of the theoretical literature treats models in which the nuclear Berry curvature is an adjustable parameter. The first-principles calculation of the nuclear Berry curvature is still in its infancy, with only a few calculations for molecular~\cite{bistoni2021} and crystalline~\cite{bonini2023} systems.

Spin-wave excitations, or magnons, constitute additional DOF in magnetic materials. These excitations are chiral from the outset; for example, in the presence of easy-axis anisotropy, spins always precess clock-wise when viewed end-on. While it is not so widely appreciated, magnon dynamics can also be formulated and computed in the context of a geometric-phase framework~\cite{niu1998,gebauer2000}.
There is no inertial term in the spin dynamics, but the Berry curvature tensor enters the EOM by describing the spin precession in response to a torque~\cite{landau1935}.
This raises the possibility of a uniform treatment of both nuclear and spin degrees of freedom in a common theoretical framework.

Crucially, the frequency (or energy) range of the magnons strongly overlaps that of the phonons. In metals these also overlap with electron-hole excitations, but we shall restrict our attention here to insulators in which there is a clear energy separation between both phonons and magnons on the one hand, and cross-gap electronic excitations on the other. In this case, both types of bosonic excitations can be treated as ``slow'' DOF, with the remainder of the electronic system following adiabatically. In such cases it is crucial to treat both phonon and spin DOF on the same footing. 

A first step was taken in this direction in Ref.~\cite{bonini2023} for the case of bulk ferromagentic CrI$_3$. In that work, a minimal model was proposed in which the nuclear Berry curvature arose entirely from the canting of Cr spins in response to atomic displacement. Though this model captured the essential physics of CrI$_3$, a general and quantitatively accurate theory should include other contributions to the Berry curvature including ``phonon-only'' Berry curvature arising from atomic displacements at fixed spin, and mixed ``spin-phonon'' Berry curvature. Such contributions can play an important role, particularly in predicting energy splittings of high-energy chiral phonons. Additionally, Ref.~\cite{bonini2023} relied on the input of experimental magnon energies, which may not always be available. 

Motivated by the need for better fundamental and quantitative understanding of phonons in TRS-broken systems, we present a generalized adiabatic treatment incorporating all relevant Hessian matrices and Berry curvatures for both phonons and spins. It should be noted that the methodology presented in this work provides a general theoretical framework for treating dynamics beyond the specific case of coupled spins and phonons. I.e., it allows for efficient \textit{ab initio} calculation of the adiabatic dynamics of any mixed set of slow variables. This goes significantly beyond the more approximate approach of Ref.~\cite{bonini2023}, which we shall refer to as the minimal spin-phonon model below. We also demonstrate an \textit{ab initio}
methodology to calculate these matrices. We conduct four case studies covering ferromagnetic (FM) and antiferromagnetic (AFM) materials in both three-dimensional (3D) and two-dimensional (2D) form. We investigate the energy splittings and symmetries of chiral phonons in these systems.

We show that, interestingly, circularly polarized chiral phonons do not always exhibit energy splittings when TRS is broken; whether this splitting occurs depends on the point group symmetry in the presence of magnetic order. Also, terms beyond the IFC are needed to correctly capture the symmetry of phonons in cases where magnetic order breaks a spatial symmetry $g$, but the combination of $g$ and time reversal $\mathcal{T}$ remains a symmetry. This limitation arises because conventional phonon calculations rely on real-symmetric IFC and atomic mass matrices that preserve TRS.%
\footnote{If $g$ is a symmetry of the nonmagnetic crystal, but neither $g$ nor $g\mathcal{T}$ remains a symmetry in the presence of magnetic order, the symmetry breaking of $g$ will still manifest itself in the real IFC matrix.} We show that correct accounting for phonon symmetries is crucial for determining Raman/IR activities.

This paper is organized as follows. In \sref{theory}, we establish the theoretical framework via the Lagrangian formalism of adiabatic dynamics, and derive EOM that treat spins and phonons on an equal footing. We also revisit phonon angular momentum and introduce the concept of atom-resolved phonon angular momentum. \Sref{met} details the proposed method for calculating all the matrices involved in the present approach, along with the computational details. In \sref{res}, we present our results, which include the analysis of phonons and magnons in 3D FM CrI$_3$ bulk (\sref{res-cri3}), 3D AFM Cr$_2$O$_3$ bulk (\sref{afm-3d}), and 2D systems (\sref{res-2d}) such as FM CrI$_3$ monolayer (\sref{cri3-2d}) and AFM VPSe$_3$ monolayer (\sref{afm-2d}). \Sref{diss} provides a discussion of the role of spin-orbit coupling (\sref{soc}) and future experimental investigations (\sref{out}). Finally, we summarize our findings and present concluding remarks in \sref{conc}.

%=================================================
\section{Theoretical background}
\seclab{theory}
%=================================================

We restrict our considerations to the case of insulating magnetic materials in which there is a clear separation between the energy scales of the phonons and the cross-gap electronic excitations. The magnetic order insures that there will also be dynamics associated with spin fluctuations, i.e., the magnons.  If the magnon frequencies would be higher than, and robustly gapped from, those of the phonons, it would be possible to treat all electronic excitations, including the magnons, as adiabatically following the phonon DOF.  However, this is almost never the case.  In the present work, we therefore treat both the lattice and spin DOF on a similar footing, while assuming that there is still a large energy gap between the top of the phonon or magnon spectrum and the onset of cross-gap electronic excitations.

We formulate our theory in the context of a first-principles mean-field theory such as DFT, where the dynamics of the nuclei is treated classically while the electronic system is evolved according to the time-dependent Shr\"{o}dinger equation.  This is essentially the domain of time-dependent DFT (TDDFT),
but here we aim to treat the spin DOF as slow semiclassical variables alongside the nuclear displacements.  This requires a separation of the electronic DOF into a small number of spin DOF and the remaining large number of electronic excitations on the scale of the band gap or above.

To do so, we define a ``spin'' unit vector on a magnetic ion to be the direction of the average spin density inside a Wigner-Seitz sphere centered at that site, and subsequent calculations of electronic ground states and energies are always computed with these spin variables constrained. 
The essential requirement is that the remaining electronic system, so constrained, should be free of any remaining slow DOF, i.e., any below-band-gap excitations. The implementation of the Wigner-Seitz sphere constraint is not a serious obstacle in practice, as most DFT code packages have features for carrying out electronic minimizations under the constraint of fixed spin orientations defined in this way. We emphasize that we treat not only the lattice displacements, but also the spin cantings, in a harmonic approximation about the ground-state reference structure. We treat only collinear easy-axis systems here, and assume that the spin cantings with respect to this axis are small.

To finish a discussion of the approximations of our theory, we note that we use ``ordinary'' adiabatic dynamics, in which the adiabatic perturbation theory is carried only to first order in the rate of change of nuclear or spin variables.  This is well justified as long as the gap separating phonons and magnons from interband electronic excitations is large. And finally, we shall shortly make a harmonic approximation, in which the atomic displacements and spin cantings are expanded to leading order around a ground-state reference configuration.

With these understandings, we turn now to a detailed presentation of our methodology. 

%=================================================
\subsection{Lagrangian formulation of adiabatic dynamics}
\seclab{lag-form}
%=================================================

While the Hamiltonian formalism is commonly used in the literature to analyze chiral phonons~\cite{mead1979,zhang2014,bistoni2021,saparov2022,bonini2023}, we will instead start with the Lagrangian formalism. We show in \sref{model-full} that this approach is well-suited to developing a comprehensive model of the coupled spin-phonon dynamics, while avoiding the difficulties associated with defining canonical momentum for spins. The Lagrangian takes the form 
\begin{equation}
L = \frac{1}{2} \sum_i M_i \dot{Q}_i^2 - \epsilon(Q) + \hbar \sum_i \dot{Q}_i A_i \,,
\eqlab{lag-general}
\end{equation}
where the configuration $Q$ can represent any slow variables.  While the formulation is general, we shall focus on the case that $Q$ represents both the nuclear coordinates and the spin variables, where the latter act as constraints on the spin moments as explained above. For clarity of presentation, we assume a finite number of nuclear and spin DOF, as for a molecule or the $\Gamma$-point modes of a periodic crystal.

The first term in \eq{lag-general} is the kinetic energy associated with the $i$-th degree of freedom, where $M_i$ is the nuclear mass for the phonon variables or zero for the spin-canting variables.  The second term is the potential energy, and the third represents the coupling between the time derivative of the adiabatic variable $Q_i$ and the Berry potential $A_i$~\cite{niu1998}. The latter is defined as
\begin{equation}
    A_i(Q) = \me{\psi(Q)} {i\frac{\partial}{\partial Q_i}} {\psi(Q)} \,,
    \eqlab{a-def}
\end{equation}
where $\ket{\psi(Q)}$ represents the electronic wave function at constrained nuclear coordinates and spin orientations $Q$. 

By solving the Euler-Lagrangian equation for the Lagrangian defined in \eq{lag-general}, we obtain
\begin{equation}
    M_i \ddot{Q}_i + \hbar \dot{A}_i =  -\partial_i \epsilon(Q) + \hbar \sum_j \dot{Q}_j \partial_i A_j \,,
    \eqlab{eom-lang-1}
\end{equation}
where $\partial_i$ denotes $\partial/\partial Q_i$. By using $\dot{A}_i = \sum_j \partial_j A_i \dot{Q}_j$, we can simplify \eq{eom-lang-1} to
\begin{align}
    M_i \ddot{Q}_i  &=  -\partial_i \epsilon(Q) + \hbar \sum_j \dot{Q}_j (\partial_i A_j - \partial_j A_i) \,, \nn
    &= -\partial_i \epsilon(Q) + \sum_j G_{ij}(Q) \dot{Q}_j \,.
    \eqlab{eom-lang}
\end{align}
Here 
\begin{equation}
   G_{ij}(Q) = \hbar \Omega_{ij}(Q) = \hbar (\partial_i A_j - \partial_j A_i) \,,
   \eqlab{G-matrix}
\end{equation}
where $\Omega_{ij}$ is the Berry curvature, and is therefore gauge-invariant. Although the gauge-dependent quantity $A_i(Q)$ appears in the Lagrangian, the EOM are gauge-invariant since only $G_{ij}(Q)$ appears.

In this work, we focus on small oscillations near equilibrium. To simplify the analysis, we introduce the generalized displacement vector $q$ defined via $Q_i = Q^{(0)}_i + q_i$, where $Q^{(0)}_i$ is the equilibrium value of the $i$-th degree of freedom.
Introducing the harmonic approximation, we expand the potential energy $\epsilon(Q)$ in terms of $q$ as 
\begin{equation}
    \epsilon(Q) = \epsilon(Q_0) + \frac{1}{2} \sum_{ij} K_{ij} q_i q_j + ... \,,
    \eqlab{eps-expansion}
\end{equation}
where $K_{ij}$ is the Hessian matrix in $q_i$ and $q_j$, i.e., $K_{ij} = \partial_i \partial_j \epsilon(Q)|_{Q=Q_0}$.
The EOM for $q$ is then given by 
\begin{equation}
    M_i \ddot{q}_i = - \sum_j K_{ij} q_j + \sum_j G_{ij} \dot{q}_j  \,,
    \eqlab{eom-general}
\end{equation}
where $G = G(Q)|_{Q=Q_0}$ is also computed at the reference configuration $Q_0$.
Conventional treatments of phonons in isolation typically use only mass and force-constant matrices $M$ and $K$, while spin dynamics in isolation is described by the anisotropy tensor $K$ and Berry curvature $G$. 
Note that $M$ is real diagonal, $K$ is real symmetric, and $G$ is real antisymmetric. 

To determine the frequencies, we substitute $q_i(t) = e^{-i\omega t} q_i$ into \eq{eom-general}, yielding 
\begin{equation}
     -\omega_n^2 M \ket{q_n} = -K \ket{q_n} - i\omega_n G \ket{q_n} \,,
    \eqlab{eom-harm}
\end{equation}
where $\ket{q_n}$ is a column vector with the $i$-th component $q_{n,i}$ corresponding to the $n$-th mode associated with DOF $i$. \Eq{eom-harm} is easily solved using, e.g., the methods of \sref{met-solve}. The above treatment provides a semiclassical theory of the adiabatic dynamics of the system.

%=================================================
\subsection{Mead-Truhlar approach without explicit spin degrees of freedom}
\seclab{model-mt}
%=================================================

If we limit the slow variables $Q$ to include only atomic coordinates, i.e., allowing all electronic degrees of freedom (including spins) to be in their instantaneous ground state for a given $Q$, we restore the treatment of Mead and Truhlar in Ref.~\cite{mead1979}. To indicate the specialization to atomic coordinates, in this section we replace $Q$ and $q$ by $R$ and $u$, where $R$ denotes the equilibrium position for atomic coordinates, and $u$ denotes the atomic displacement from the equilibrium. Now the EOM for $u$ in the harmonic approximation is 
\begin{equation}
    M^{\rm MT}_l \ddot{u}_l = - \sum_n K^{\rm MT}_{lm} u_m + \sum_n G^{\rm MT}_{lm} \dot{u}_m  \,,
    \eqlab{eom-mt}
\end{equation}
where $l$ and $m$ are composite indices for $I\alpha$, $I$ runs over atoms and $\alpha$ represents a Cartesian direction. \Eq{eom-mt} shows that $G^{\rm MT}_{mn} \dot{u}_n$ corresponds to a force acting on coordinate $m$ that is proportional to the velocity of coordinate $n$. An alternative derivation of \eq{eom-mt} using the quantum theory in the Hamiltonian framework is given in Appendix~\ref{ap:ham}. 

The conventional treatment of phonons~\cite{gonze1997,baroni2001} is recovered by discarding the term involving the nuclear Berry curvature $G^{\rm MT}$ in \eq{eom-mt}. This is justified in TR-invariant systems, where $G^{\rm MT}$ vanishes by symmetry. However, $G^{\rm MT}$ is often neglected even when the system is not TR symmetric, with the consequences that \eq{eom-mt} obeys TRS and the phonons will not have the correct symmetry of the TRS-broken system. This will force modes with opposite chirality to be degenerate at the zone center, which is not necessarily the case in a magnetic material.

Recent works have used \eq{eom-mt} to demonstrate splitting of chiral modes as a result of TRS breaking~\cite{saparov2022,bonini2023}. However, it was demonstrated in Ref.~\cite{bonini2023} that for CrI$_3$, the main contribution to $G^{\rm MT}$ comes from canting and precession of spins. As mentioned earlier, the energy scale for spin rotations corresponds to the frequency of the magnons, which is close to that of the phonons in most systems. Consequently, the assumption that atomic displacements are the only slow DOF in the system, which led to \eq{eom-mt}, is not valid.  Reference~\cite{bonini2023} developed a Hamiltonian formalism for coupled spin-phonon dynamics that we refer to here as the ``minimal spin-phonon model." In the next section we will develop a more general Lagrangian-based approach that is well suited to treating phonon and spin dynamics together.

%=================================================
\subsection{Treatment of spins and phonons on the same footing}
\seclab{model-full}
%=================================================

We now return to the framework of \eq{eom-general} in which $q_i$ includes both nuclear and spin DOF. The Euler-Lagrange EOM derived from \eq{eom-general} is
\begin{equation}
    \sum_j M_{ij} \ddot{q}_j = -\sum_j K_{ij} q_j + G_{ij} \dot{q}_j\,,
    \eqlab{eom-full}
\end{equation}
where $i$ runs over both phonon and spin DOF. 
In the phonon sector, $q_i=u_{I\alpha}$ is a shorthand for a small displacement of atom $I$ in Cartesian direction $\alpha=\{x,y,z\}$.
In the spin sector, $q_i$ denotes a small spin canting $q_i=s_{J\beta}$, where $J$ runs only over magnetic ions and $\beta$ indexes the spin canting in the two directions orthogonal to the ground-state spin orientation. Specializing to easy-axis systems with spin axis along $\pm\hat{z}$, we let $\beta$ run over only the two in-plane Cartesian directions. 
That is, $q_i = s_{J \beta}=S_{J\beta}/|S_J|$ describes the component $\beta=\{x,y\}$ of the unit vector of spin $S_J$ located on the $J$-th magnetic ion.
 
The matrix $M_{ij}$ in \eq{eom-full} is the diagonal mass matrix introduced in \eq{lag-general}, with zero entries for the spin DOF; $K_{ij} = \partial_i \partial_j \epsilon(\q)$ is a generalized Hessian matrix; 
and $G_{ij} = \hbar \Omega_{ij} = \hbar (\partial_i A_j - \partial_j A_i)$ embodies the Berry curvature as in \eq{G-matrix}.  In these last two expressions, $\partial_j$ denotes $\partial/\partial q_j$, which can be a derivative with respect to either nuclear or spin-canting coordinates, evaluated at the reference ground-state configuration. 
It should be emphasized that in our formulation, the Hessian matrices are expanded to quadratic order in phonon or magnon amplitudes, resulting in a harmonic theory with infinite lifetimes for both phonons and magnons. This is also evident in the hermiticity of \eq{eom-full}. While including anharmonic interactions beyond quadratic order would allow decay of a phonon or magnon excitation into two or more lower-energy excitations, thereby rendering their lifetimes finite, such effects are not considered in this work and remain an area for future investigation.

In this context, derivatives with respect to nuclear coordinates must be taken at fixed spin. That is, the matrices $K$ and $G$ are now computed in terms of the electronic quantum state $\ket{\psi(R,s)}$, rather than $\ket{\psi(R)}$ as in \sref{model-mt}. As explained earlier, this requires a calculation of the electronic ground state subject to constrained spin orientations. While there is some freedom in the definition of the spin unit vector, we follow the established approach of defining it in terms of the integrated spin density inside a Wigner-Seitz sphere, as discussed in \sref{met-com}. This choice encodes the distinction between ``spin'' and ``other electronic'' DOF in our theory. 

To simplify the analysis, we can partition all DOF $i$ into phonon DOF (labeled as $\mathrm{p}$) and spin DOF (labeled as $\mathrm{s}$). The matrices in \eq{eom-full} can then be represented using a block structure as 
\begin{align}
M =& \begin{pmatrix}
    \Mp & 0 \\ 
    0 & 0
    \end{pmatrix} \,, \nn
K =& 
\begin{pmatrix}
    \kpp & \kps \\
    \ksp & \kss 
\end{pmatrix} \,, \nn 
G =& 
\begin{pmatrix}
    \gpp & \gps \\
    \gsp & \gss 
\end{pmatrix} \,.
\eqlab{full-mat}
\end{align}
We use the term ``bare phonons" to refer to phonons that are calculated without considering $\gpp$, $\gps$ ($\gsp$), or $\kps$ ($\ksp$), while phonons calculated with the inclusion of those terms are referred to as ``perturbed phonons." The term ``perturbed" is used in recognition of the fact that the influence of these terms is generally small, although we solve \eq{eom-full} exactly. Nevertheless, we also provide a perturbation analysis of $\ksp$, $\gsp$, and $\gpp$ in Appendix~\ref{ap:pert}, illuminating the physical implications of each term, and thereby offering valuable insight.

It is worth noting that $\gpp$, $\gps$ ($\gsp$), and $\kps$ ($\ksp$) are zero in the absence of spin-orbit coupling (SOC) in collinear systems, a point elaborated upon in \sref{soc} below. In that case, the broken TRS in the spin sector is never communicated to the orbital electronic sector or, in turn, to the phonon sector. 

Importantly, in the case that $\kps$ ($\ksp$) and $\gps$ ($\gsp$) vanish, phonons and spins decouple, and \eq{eom-full} reduces to the EOM for phonons and magnons separately. In the phonon sector, it reduces to \eq{eom-mt} which corresponds to the approach proposed by Mead and Truhlar~\cite{mead1979}. Meanwhile, in the magnon sector, it reduces to 
\begin{equation}
    \hbar \oss \ket{\dot{s}} = \kss \ket{s} \,,
    \eqlab{eom-magnon}
\end{equation}
which aligns with the EOM for magnons presented in Ref.~\cite{niu1998}. 
Additionally, when $\oss$ is simplified to consider only isolated spinors, rendering inter-spin elements negligible, \eq{eom-magnon} reduces to the well-known Landau-Lifshitz equation~\cite{landau1935,niu1999}. 
Further, if the energies of phonons are considerably lower than those of magnons, \eq{eom-full} in the phonon sector also reduces to the Mead-Truhlar approach, as discussed in Ref.~\cite{bonini2023}. 

The present approach defined by \eqs{eom-full}{full-mat} uses Hessian matrices and Berry curvature tensors in terms of all DOF. In contrast, the minimal spin-phonon model presented in Ref.~\cite{bonini2023} included only the spin-spin component of the Berry curvature, so that $\gpp$, $\gps$, and $\gsp$ were assumed to vanish. Furthermore, the minimal spin-phonon model in Ref.~\cite{bonini2023} only includes one bare phonon doublet and one bare magnon. In Appendix~\ref{ap:sp-model}, we will discuss a more comprehensive version of the spin-phonon model which incorporates all bare phonons and magnons but still neglects $\gpp$, $\gps$, and $\gsp$, and discuss its connection to the minimal spin-phonon model and its equivalence with the widely employed Landau-Lifshitz equation~\cite{landau1935}. 
We demonstrate that our approach, through the inclusion of non-trivial $\opp$, $\ksp$, and $\osp$ terms, provides a fruitful generalization of the Landau-Lifshitz equation.

%=================================================
\subsection{Phonon angular momentum}
\seclab{model-pam}
%=================================================

Before discussing the first-principles methodology to calculate the terms in \eq{eom-full} and \eq{full-mat}, we review the definition of phonon angular momentum and introduce the concept of atom-resolved phonon angular momentum, as this will be important for characterizing the chiral modes in \sref{res}. The definition of phonon angular momentum can be found in the literature~\cite{zhang2014}, and we briefly revisit the relevant definitions here. First, we note that by solving \eq{eom-full}, one can obtain an energy eigenvalue $\omega_n$ and the corresponding mode with both phonon and spin components, i.e.,
$\ket{q_n} = \ket{u_n} \oplus \ket{s_n}$, where $n$ runs over different solutions of \eq{eom-full}.
Since we are mainly interested in the phonon sector in the present work, we define the atom-resolved phonon angular momentum using the phonon part $\ket{u_n}$ of the mode. 

The solutions to the EOM described by \eq{eom-full} yield energies that differ from those of the bare phonons. However, in all of the systems examined in this work, the differences between these energies and those of the bare phonons are relatively small. As such, we can identify the mode $\ket{q_n}$ as``phonon-like" if its energy $\omega_n$ is in close proximity to that of a bare phonon. Furthermore, we note that for a phonon-like mode $\ket{q_n}$, the $\ket{s_n}$ contributions are very small compared to $\ket{u_n}$. In systems where the frequencies of the phonons and magnons coincide, the aforementioned conditions may not be met; nonetheless, in all cases we consider in this work, the zone-center phonons and magnons of relevance exhibit distinct energies, thus ensuring well-defined phonon-like and magnon-like modes. 

For phonon-like modes, we continue to adopt the normalization convention
\begin{equation}
    \me{u_n}{\Mp}{u_n} = 1 \,,
\end{equation}
even though $\me{u_m}{\Mp}{u_n}$ is no longer exactly zero for $m \neq n$.
It is also possible to have a ``perturbed magnon" solution, which is a magnon-like solution with tiny phonon components. 

The definition of phonon angular momentum was originally proposed in Ref.~\cite{zhang2014}. For a phonon-like mode $\ket{u_n}$, the atom-resolved phonon angular momentum (ARPAM) $L_{n, Iz}$, for atom  $I$ of mass $M_I$ in the $z$ direction is 
\begin{align}
    L_{n, Iz} &= \hbar M_I
    \begin{pmatrix}
        u_{n, I x}^* \; u_{n, I y}^*
    \end{pmatrix}
    \begin{pmatrix}
        0 & -i \\
        i & 0
    \end{pmatrix}
    \begin{pmatrix}
        u_{n, I x} \\
        u_{n, I y}
    \end{pmatrix} \nn
    & = 2 \hbar M_I \Im[u_{n, I x}^* u_{n, I y}] \,,
    \eqlab{lz-atom}
\end{align}
with $L_{n, Ix}$ and $L_{n, Iy}$ defined similarly by cyclic permutation of Cartesian indices. Note that if the projection of a mode vector on a given atom is of the form $\hat{x}+i\hat{y}$, the atom undergoes a counterclockwise rotation when viewed from above and contributes a positive $L_z$. 
The total phonon angular momentum (PAM)%
\footnote{We are aware that in some literature, the PAM is referred to as ``pseudo angular momentum"~\cite{zhang2015}, which is a non-zero integer multiple of $\hbar$ if the phonon respects $C_n$ symmetry with a nontrivial eigenvalue. However, it is important to note that the phonon angular momentum discussed in this paper is the kinetic angular momentum and is not conserved in the absence of infinitesimal rotational symmetry in the lattice. In contrast, the pseudo angular momentum is conserved up to $n\hbar$ if the system possesses $C_n$ symmetry. In \sref{res}, we refer to the $C_3$ eigenvalue as $\chi(C_3)$.}
$L_{n, z}$ is defined as 
\begin{equation}
L_{n, z} = \sum_I L_{n, Iz}\,,
\eqlab{lz-total}
\end{equation}
which is the sum of the angular momenta $L_{n, Iz}$ over all atoms $I$. 

The bare phonons solve the secular equation involving only $\Mp$ and $\kpp$, which is equivalent to the conventional phonon treatment of \eq{eom-mt} with $G=0$. These bare phonons can always be chosen real, in which case it follows from \eq{lz-atom} that the full PAM and individual ARPAM always vanish. In the case of degenerate modes it may be possible to choose chiral linear combinations, but the trace over the degenerate set of modes always results in zero PAM and ARPAM. This is a consequence of the fact that TRS has not yet been broken at the bare level of description.

%=================================================
\section{Methods}
\seclab{met}
%=================================================

%=================================================
\subsection{Finite difference method}
\seclab{met-fdm}
%=================================================

This section demonstrates how to compute all matrices in \eq{full-mat} using finite-difference methods in the context of first-principles calculations. The nuclear masses $\Mp$ are trivially known.
All calculations are carried out using the DFT methodology described in \sref{met-com} at fixed atomic coordinates and fixed spin orientations, where the latter are defined in terms of an integration of the spin density over a Wigner-Seitz sphere as mentioned earlier. 

The force-constant matrix $\kpp$, which is also the Hessian matrix of the energy with respect to nuclear DOF, is defined as
\begin{equation}
    \kpp_{lm} = \frac{\partial^2 \epsilon}{\partial u_l \partial u_m} = -\frac{\partial F_l}{\partial u_m} \,.
    \eqlab{k-pp-force}
\end{equation}
Here $l$ and $m$ run over nuclear DOF and $\epsilon$ is the total energy $\epsilon(s,R)$ of the configuration $s,R$. The matrix element $\kpp_{lm}$ is computed by taking the finite difference of the Hellmann-Feynman forces $F_l$~\cite{gonze1997,baroni2001} while constraining the spin directions to lie along $\hat{z}$.

The Hessian matrix of the energy with respect to spin DOF is denoted by $\kss$, with elements defined as
\begin{equation}
    \kss_{a b} = \frac{\partial^2 \epsilon}{\partial s_a \partial s_b} \,,
\end{equation}
where $a$ and $b$ run over spin DOF. To calculate $\kss$, we compute the second derivatives of the total energy with respect to small canting of the spins.
Specifically, letting $s_a=s_{I\alpha}$ and $s_b=s_{J\beta}$, for each pair $(I,J)$ we compute the energies of $(\alpha,\beta)$ = $\pm(0.02, 0)$, $\pm(0,0.02)$, and $\pm(0.02,0.02)$ relative to the ground state,
while constraining all other spin moments to remain along $\hat{z}$. 

The spin-phonon Hessian matrix $\ksp$ is defined as
\begin{equation}
    \ksp_{a l} = \frac{\partial^2 \epsilon}{\partial s_a \partial u_l} = -\frac{\partial F_l}{\partial s_a} \,,
    \eqlab{k-sp-force}
\end{equation}
where $a$ and $l$ respectively run over the spin and nuclear DOF. $\ksp_{a l}$ is computed by taking the finite-difference derivative of the force $F_l$ with respect to the spin coordinate $s_a$.

An alternative approach to calculating $\ksp$ was described in Ref.~\cite{bonini2023}. Near the ground state, the energy $\epsilon$ of the entire system can be expanded as
\begin{equation}
    \epsilon = \frac{1}{2} \sum_{l m} \kpp_{l m} u_l u_m + \sum_{a l} \ksp_{a l} s_a u_l + \frac{1}{2} \sum_{a b} \kss_{a b} s_a s_b \,.
\end{equation}
As $\epsilon$ is minimized with $s_a$, we have
\begin{equation}
    \frac{\partial \epsilon}{\partial s_a} = \sum_{l} \ksp_{a l} u_l + \sum_{b} \kss_{a b} s_b = 0 \,.
\end{equation}
We define a spin response matrix $\chi^{\rm (sp)}$ as
\begin{equation}
    \chi^{\rm (sp)}_{a l} = \frac{\partial s_a}{\partial u_l} \simeq \frac{s_a}{u_l} \,,
    \eqlab{chi}
\end{equation} 
where the second equality holds if both $s_a$ and $u_l$ are small. Then we can obtain
\begin{equation}
    \ksp = -\kss \chi^{\rm (sp)} \,,
    \eqlab{ksp-chi}
\end{equation}
where we have restored the matrix form for simplicity. In practice, one can perturb the ground state structure with $u_l$ by manually moving the atoms from the equilibrium position and calculate the spin canting $s_a$ with respect to $u_l$. We used \eq{ksp-chi} to calculate $\ksp$ for bulk CrI$_3$, and the result is consistent with \eq{k-sp-force}. However, due to the slow convergence of spin relaxation, we recommend using \eq{k-sp-force} to calculate $\ksp$.

In \eq{full-mat}, $G$ is simply $\hbar$ times $\Omega$, where $\Omega$ is the Berry curvature. 
The latter is computed using Stokes' theorem as expressed by 
\begin{equation}
    \Omega_{ij} = \frac{\Phi_{ij}}{2 \abs{\delta q_i \wedge \delta q_j}} \,,
    \eqlab{omega-comp}
\end{equation}
where $i$ and $j$ run over all DOF, and $\Phi_{ij}$ is the Berry phase around a diamond-shaped region of parameter space whose area appears in the denominator. Specifically for, e.g., a finite system with electronic ground state wave function $\ket{\psi}$,
\begin{align}
    \Phi_{ij} = -\Im \ln 
    [\ip{\psi(+\delta q_i)}{\psi(+\delta q_j)} 
    \ip{\psi(+\delta q_j)}{\psi(-\delta q_i)} \nn
    \ip{\psi(-\delta q_i)}{\psi(-\delta q_j)}
    \ip{\psi(-\delta q_j)}{\psi(+\delta q_i)} ] \,.
    \eqlab{phi-mole}
\end{align}

In an extended crystal with a single occupied band, one must sum over the Bloch wave vector $k$ in the Brillouin zone to obtain $\Phi_{ij}=N_k^{-1}\sum_k\Phi_{ij}^{(k)}$, where $\Phi_{ij}^{(k)}$ is defined as in \eq{phi-mole} but with $\psi$ replaced by the Bloch function $\psi_k$.
$\Phi_{ij}$ now has the interpretation of a Berry phase per unit cell, consistent with the interpretation of the Hessian $K$ as an energy per unit cell.
To extend $\Phi_{ij}^{(k)}$ to the multiband case, we can replace the inner product of two Bloch states with the overlap matrix in the usual way~\cite{vanderbilt_book} as 
\begin{align}
    \Phi_{ij}^{(k)} =  - \Im \ln \det
    [ M^k(+\delta q_i, +\delta q_j) M^k(+\delta q_j, -\delta q_i) \nn
    M^k(-\delta q_i, -\delta q_j) M^k(-\delta q_j, +\delta q_i) ]\,.
    \eqlab{phi-mb}
\end{align}
Here, the overlap matrices are defined as 
\begin{equation}
    M^k_{mn}(\delta q_i,\delta q_j) = \ip{\psi_{mk}(q_i)}{\psi_{nk}(q_j)} \,,
    \eqlab{overlap}
\end{equation}
where $m$ and $n$ are band indices. 
When computing the matrices, a finite difference of $0.015$\,\AA\ is used for the phonon DOF, and $0.02$ is used for the spin DOF. 
We take care to choose these finite differences to ensure they remain within the linear regime.
The numerical values of $\kss$ and $\gss$ for all four studied materials can be found in Appendix~\ref{ap:kss}.

%=================================================
\subsection{First principles calculations}
\seclab{met-com}
%=================================================

In this section, we provide computational details for calculating matrices using the finite difference method described in \sref{met-fdm}. The reported DFT calculations are performed using the Vienna Ab-initio Simulation Package (VASP)~\cite{kresse1993,kresse1996,kresse1999}, employing the local-density-approximation (LDA) exchange-correlation functional~\cite{perdew1981} and the projector-augmented wave~\cite{blochl1994} method, with Cr $3s^2 3p^6 3d^5 4s^1$, I $5s^2 5p^5$, O $2s^2 2p^4$, V $3s^2 3p^6 3d^5$, P $3s^2 3p^3$, Se $4s^2 4p^4$ pseudopotential valence configurations. A plane-wave cutoff of 520\,eV is adopted for the CrI$_3$ calculations, and 500\,eV for the Cr$_2$O$_3$ and VPSe$_3$ systems. 

All structures are relaxed using the local spin-density approximation (LSDA), and the convergence criteria for forces and energies are $10^{-3}\,\mbox{eV}/\mbox{\AA}$ and $10^{-8}$\,eV, respectively. After relaxation, the wave functions are calculated using static calculations with a convergence criterion of $10^{-10}$\,eV for energies. Spin-orbit coupling, which is essential to the physics described in this work, is included in all static calculations except structural relaxations.

We use $\Gamma$-centered Monkhorst-Pack $k$-points meshes~\cite{monkhorst1976} for all calculations,
specifically
$5\!\times\!5\!\times\!5$ and
$7\!\times\!7\!\times\!7$ for bulk CrI$_3$ and Cr$_2$O$_3$ respectively, and
$7\!\times\!7\!\times\!1$ and
$6\!\times\!6\!\times\!1$ for monolayer CrI$_3$ and VPSe$_3$ respectively.
For Cr$_2$O$_3$ and VPSe$_3$, the Dudarev-type DFT+U approach~\cite{dudarev1998} is used,
with values of $U$=4.0\,eV and $J$=0.6\,eV for Cr and $U$=3.25\,eV and $J$=0 for V
(adapted from Refs.~\cite{mu2019,chittari2016}). For constrained local-moment calculations, the Wigner-Seitz radii for Cr and V are 1.164 and 1.217\,\AA, respectively. 
The overlap matrix (\eq{overlap}) is calculated as in Ref.~\cite{turiansky2021}. Symmetry analysis is performed using the FINDSYM~\cite{stokes2005} and spglib~\cite{togo2018} packages, while figures are rendered using VESTA~\cite{momma2011}.

%=================================================
\subsection{Solution of the equations of motion}
\seclab{met-solve}
%=================================================

In this section, we introduce a practical method for solving the EOM in \eq{eom-full}. Given that the generalized mass matrix is non-invertible, conventional approaches are not directly applicable. We proceed as follows, although other potential methods may be suitable. 

First, we rewrite \eq{eom-full} as 
\begin{equation}
    \frac{d}{dt} 
    \begin{pmatrix}
        u \\
        \dot{u} \\
        s
    \end{pmatrix} = 
    \begin{pmatrix}
        0 & 1 & 0 \\
        A_1 & A_2 & A_3 \\
        B_1 & B_2 & B_3 
    \end{pmatrix}
    \begin{pmatrix}
        u \\
        \dot{u} \\
        s
    \end{pmatrix} \,,
    \eqlab{eom-linemat}
\end{equation}
which can be shown to reduce to \eq{eom-full} with the definitions
\begin{align}
    A_1 &= -[\Mp]^{-1} \kpp + [\Mp]^{-1} \gps [\gss]^{-1} \ksp \,,\nn
    A_2 &=  [\Mp]^{-1} \gpp - [\Mp]^{-1} \gps [\gss]^{-1} \gsp \,, \nn
    A_3 &= -[\Mp]^{-1} \kps + [\Mp]^{-1} \gps [\gss]^{-1} \kss \,, \nn
    B_1 &=  [\gss]^{-1} \ksp \,, \nn
    B_2 &= -[\gss]^{-1} \gsp \,, \nn
    B_3 &=  [\gss]^{-1} \kss \,.
\end{align}
The eigenvalues (multiplied by $i$) of the matrix in \eq{eom-linemat}, denoted as $\omega_n$, correspond to the solutions of \eq{eom-full}. We restrict the index $n$ to run only over solutions for which $\omega_n > 0$, which we take to be the physically meaningful ones.

In practice we find that small numerical errors remain in the eigenvectors, which arise from the fact that a Hermitian eigensolver cannot be applied in the context of \eq{eom-linemat}. We have found that these errors can easily be removed by following with a second step in which we substitute the computed energies $\omega_n$ back into \eq{eom-full} and then apply a Hermitian eigensolver to recalculate the eigenvectors.

%=================================================
\section{Results}
\seclab{res}
%=================================================

We conduct four case studies covering both FM and AFM systems, in both 3D and 2D. 
In \sref{res-cri3}, we present the results obtained from our present approach for bulk CrI$_3$, focusing on chiral phonons. The significance of the Berry curvatures neglected in Ref.~\cite{bonini2023}, is discussed in \sref{cri3-comp}, while the relevant solutions for the magnons are discussed in \sref{cri3-magnon}.
Subsequently, in \sref{afm-3d}, we investigate the phonons and magnons in bulk Cr$_2$O$_3$ using the present approach. Shifting our focus to monolayer systems in \sref{res-2d}, we first consider 2D FM CrI$_3$ in \sref{cri3-2d}, and then 2D AFM VPSe$_3$ in \sref{afm-2d}. 

%=================================================
\subsection{Chiral phonons in bulk ferromagnetic CrI$_3$}
\seclab{res-cri3}
%=================================================

CrI$_3$ is a hexagonal van der Waals material that exhibits FM order in both bulk and monolayer phases~\cite{huang2017}. It is an insulator with inversion and three-fold rotational symmetry around the $z$-axis in both phases. The magnetic moments on the Cr ions do not break the inversion symmetry. The symmetries will be discussed in detail below. 

The crystal structure of the bulk CrI$_3$ unit cell is depicted in \fref{cri3-bulk-crystal}. For bulk CrI$_3$, the structural and magnetic symmetries are identical (space group $R\bar{3}$, point group $S_6$), regardless of the presence of FM order. 
That is, the magnetic space group is of Type I (``colorless''), in which no symmetry operations involve TR~\cite{dresselhaus2007}. 
Therefore, the zone-center phonons can be categorized into the irreducible representations (irreps) of the $S_6$ point group as $4E_g \oplus 4E_u \oplus 4A_g \oplus 4A_u$. Among these, one $A_u$ mode and one pair of $E_u$ modes are acoustic modes. In this work, we will focus on the optical modes.

\begin{figure}
\centering\includegraphics[width=0.6\columnwidth]{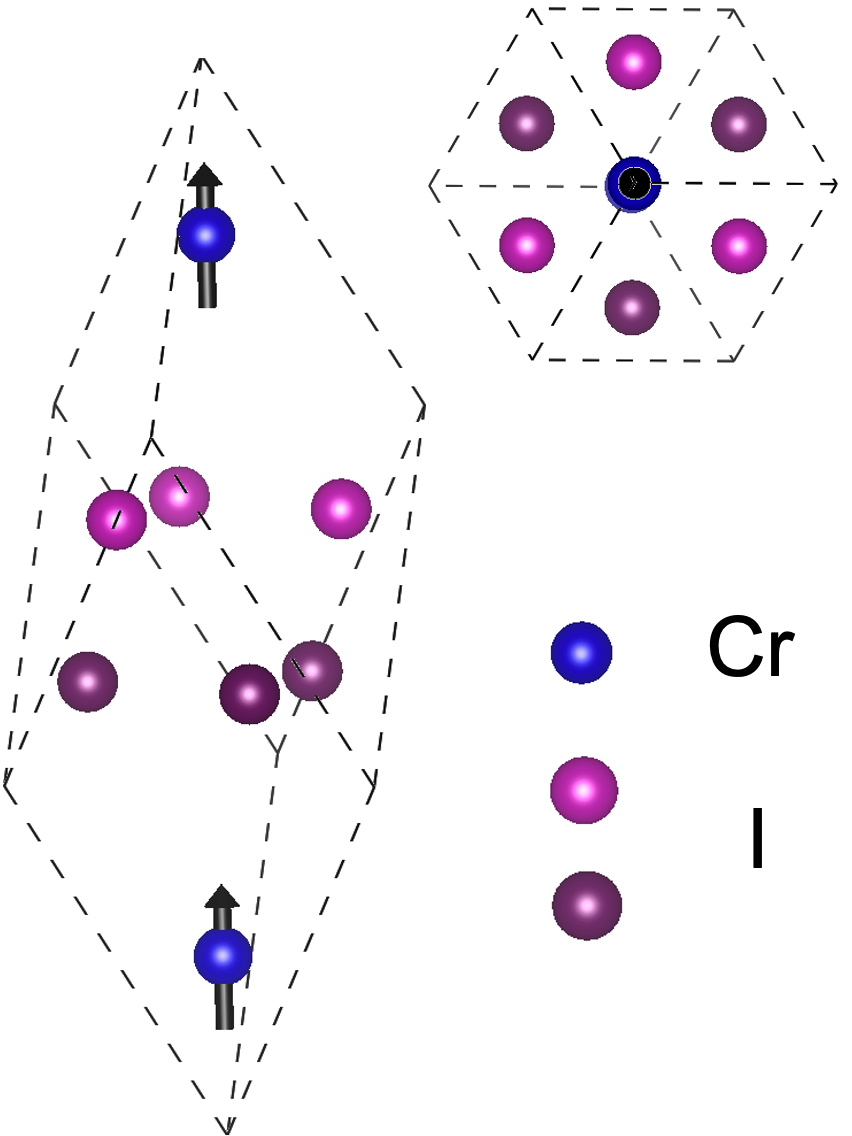}
\caption{Visualization of crystal structure and magnetic moments of bulk CrI$_3$ unit cell. Cr atoms are depicted in blue, while top-layer and bottom-layer I atoms are in bright and dark magenta, respectively. The black vectors indicate the magnetic moments, which are oriented along the $z$-direction.}
\figlab{cri3-bulk-crystal}
\end{figure}

The $E_g$ and $E_u$ irreps are complex-conjugate irreps, which are more properly decomposed further into one-dimensional irreps of opposite chirality. In the presence of TRS, or when neglecting the term involving the Berry curvature $G$ tensor in \eq{eom-mt}, the two modes making up one of these $E$ irreps are degenerate.  However, this degeneracy is broken by the FM order, which induces a non-zero Berry curvature via the SOC as discussed \sref{theory}.

\begin{table}
\caption{\label{tab:res-cri3-3d-e}
Computed properties of zone-center $E_g$ and $E_u$ phonons in bulk CrI$_3$. Energy shifts $\Delta E$ are defined relative to the bare phonon energies $E_0$, and are shown for both the full method and the spin-phonon model. $\chi(C_3)$ is the $C_3$ eigenvalue indicating the chirality of the mode ($\varepsilon=e^{2\pi i/3}$), and $L_z$ is the phonon angular momentum of \eq{lz-total}.}
\begin{ruledtabular}
\begin{tabular}{cccccc}
& bare & \multicolumn{2}{c}{present approach} & spin-phonon & \\ 
Irrep & $E_0$~(meV) & $\Delta E$~(meV) & $L_z\,(\hbar)$  & $\Delta E$~(meV) & $\chi(C_3)$ \\
\colrule
$E_g$ & $\phm 7.0000$ & $-0.0017$ & $\phm 0.1259$  & $-0.0013$ & $\varepsilon^*$
\\ 
& & $\phm 0.0022$ & $-0.1264$  & $\phm 0.0016$  & $\varepsilon\phantom{^*}$
\\
& $12.9288$ & $-0.0007$ & $-0.1325$  & $-0.0005$  & $\varepsilon^*$
\\
& & $\phm 0.0007$ & $\phm 0.1351$  & $\phm 0.0006$ & $\ve$
\\
& $13.4876$ & $-0.0007$ & $\phm 0.1740$  & $-0.0002$  & $\varepsilon^*$
\\
& & $\phm 0.0007$ & $-0.1761$  & $\phm 0.0003$  & $\ve$
\\
& $29.8518$ & $-0.0028$ & $\phm 0.8325$  & $-1.26\times10^{-6}$  & $\varepsilon^*$
\\
& & $\phm 0.0028$ & $-0.8326$  & $\phm 1.31\times 10^{-6}$  & $\ve$
\\

%\hline

$E_u$ & $10.7687$ & $-0.0052$ & $\phm 0.2352$  & $-0.0045$ & $\ve$
\\
& & $-0.0009$ & $-0.2344$  & $-0.0016$ & $\varepsilon^*$
\\
& $14.3295$ & $-0.0170$ & $\phm 0.7173$  & $-0.0175$ & $\ve$
\\
& & $-0.0042$ & $-0.7192$  & $-0.0040$ & $\varepsilon^*$
\\
& $27.8225$ & $\phm 0.0037$ & $\phm 0.9537$  & $-0.0035$ & $\varepsilon^*$
\\
& & $\phm 0.0272$ & $-0.9545$  & $\phm 0.0356$ & $\ve$
\\
\end{tabular}    
\end{ruledtabular}
\end{table}

We first present our results for the case of the $E_g$ and $E_u$ phonons in Table~\ref{tab:res-cri3-3d-e}. The energy shifts $\Delta E$ represent the differences between the bare phonon frequencies $E_0$ computed using only the force-constant term $\kpp$ in \eq{eom-mt}, and the modified energy obtained from the present approach of \eq{eom-full} or from the spin-phonon model of Appendix~\ref{ap:sp-model}.%
\footnote{The results presented in Table~\ref{tab:res-cri3-3d-e} for the spin-phonon model differ from those in Ref.~\cite{bonini2023} because in this paper, all matrices are calculated using DFT, while in Ref.~\cite{bonini2023}, experimental magnon energies are used to evaluate $\kss$, and the approximation in \eq{gss} is taken.}
The PAM $L_z$ of each phonon is also reported for the present approach.

From Table~\ref{tab:res-cri3-3d-e} we observe that all $E_g$ and $E_u$ bare phonons are doubly degenerate, while the degeneracy is broken for the perturbed phonons.  Although our numerical solution for the bare modes initially yields a pair of real phonons $\ket{u_1}$ and $\ket{u_2}$, we resolve these into eigenstates of the $C_3$ symmetry operator.  We first ensure that $\me{u_2}{C_3}{u_1}$ is positive,
flipping the sign of $\ket{u_2}$ if not, and then construct the bare chiral modes $\ket{u_\pm}=(\ket{u_1}\mp i\ket{u_2})/\sqrt{2}$
belonging to eigenvalues $\varepsilon = \exp(i 2 \pi/3)$ and $\varepsilon^*=\exp(-i2\pi/3)$ respectively.  We shall designate these as `$+$' and `$-$' modes, and refer to them as belonging to the $\varepsilon$ and $\varepsilon^*$ symmetry sectors, respectively.  The former (latter) are characterized by a clockwise (counterclockwise) rotation of the Cr atoms when viewed from above. 

The numerical solutions for the perturbed phonons automatically
generate chiral $C_3$ eigenstates, and we find that each of these is
almost identical to the bare chiral mode of the same symmetry
that is closest in energy.  There is
only a small admixture of other bare modes belonging to the same
sector.  This allows a straightforward association of bare and
perturbed modes as shown in Table ~\ref{tab:res-cri3-3d-e}.

Table~\ref{tab:res-cri3-3d-e} demonstrates that the $E_u$ phonons with $\varepsilon$ chirality have the largest energy corrections. This follows from their close energy proximity to the $E_u$ magnons, which also possess $\varepsilon$ chirality. A detailed discussion of magnons in bulk CrI$_3$ is provided in \sref{cri3-magnon}, and a perturbative treatment of phonon-magnon interactions is presented in Appendix~\ref{ap:pert}. The pronounced effect on the $E_u$ phonons is further evident in \eq{ph-en-2nd}, where the phonons with $\varepsilon$ chirality have the smallest energy denominators.

We briefly address the expected uncertainties in our first-principles calculations of the $\ksp$, $\gsp$, and $\gpp$ matrices, as these are the matrices that contribute to the energy shifts (see Appendix~\ref{ap:pert}).  First, we analyze the fitting errors for $\ksp$ associated with our finite-difference methodology, which is done via linear regression of the forces against spin canting angles. We find that the norm of the fitting error is roughly 1.5\% of the norm of $\ksp$ itself. A second useful metric is the norm of the residual resulting from symmetrization of a matrix relative to the norm of the matrix itself, which we find to be 0.4\%, 2.0\%, and 2.7\% for $\ksp$, $\gsp$, and $\gpp$ respectively. The perturbation analysis of \eqs{ph-en-2nd}{gpp-omega} indicates that the energy shifts and splittings are quadratic in $\ksp$ and $\gsp$ and linear in $\gpp$. This implies overall errors on the order of 3-4\% in the values reported above for these shifts and splittings.

% \color{black}

In \fref{vis-eg}, we visualize the real and imaginary parts of
two $E_g$ chiral phonons around 7~meV, denoted as $E_g^{(1)}$
for the mode with lower energy and $E_g^{(2)}$ for the mode with
higher energy. It is apparent that those two chiral phonons are
approximately complex conjugates of each other. 

\begin{figure}
\centering\includegraphics[width=\columnwidth]{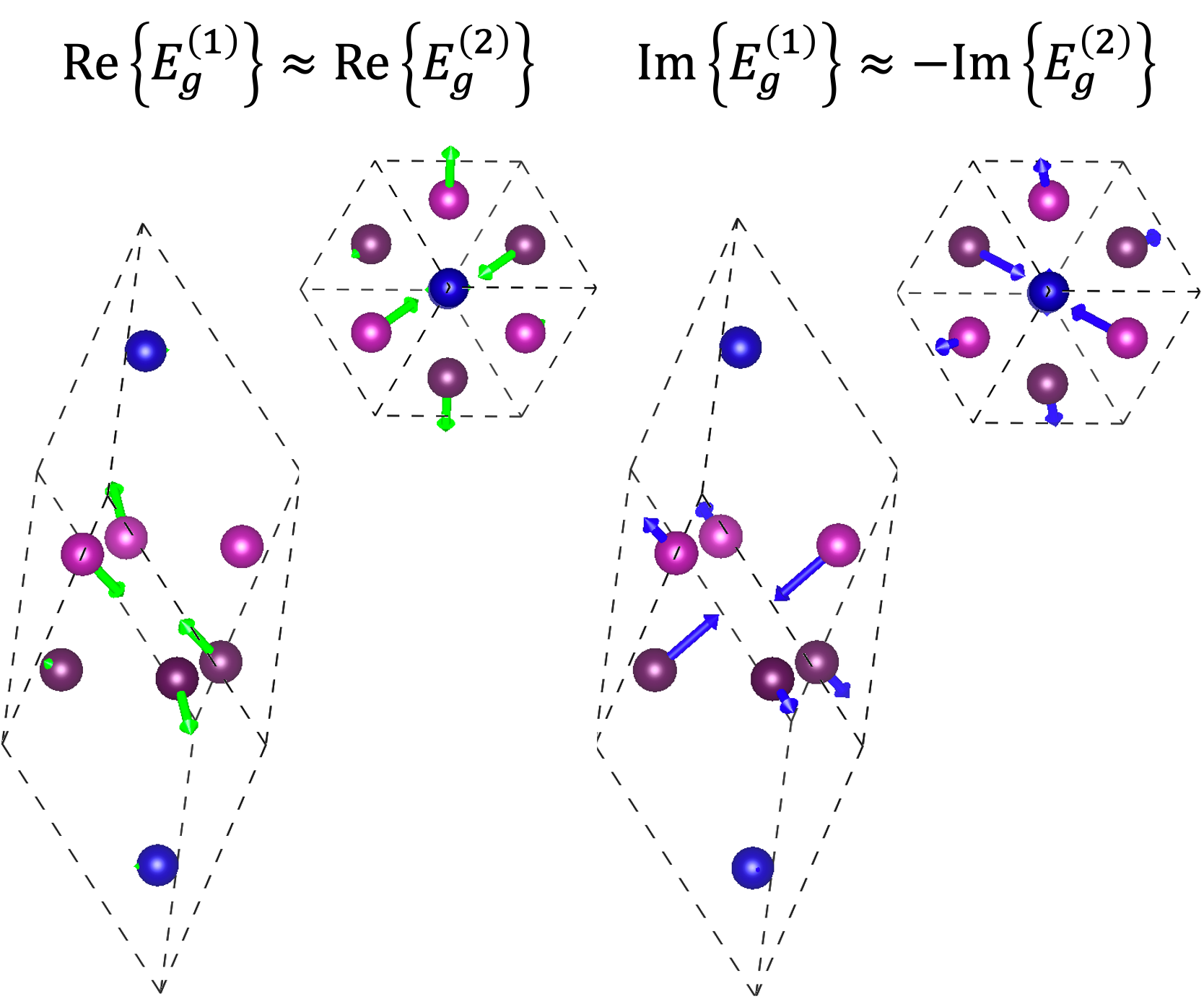}
\caption{Visualization of the real and imaginary components of the chiral phonons $E_g^{(1)}$ and $E_g^{(2)}$. Cr atoms are depicted in blue, while top-layer and bottom-layer I atoms are in bright and dark magenta, respectively. The green vectors represent the real part of the phonon displacement, and the blue vectors represent the imaginary part. The real part of $E_g^{(1)}$ and $E_g^{(2)}$ are nearly identical, while the imaginary parts are in opposite directions.}
\figlab{vis-eg}
\end{figure}

Note that the argument about energy splitting based on point group irreps can be generalized to any magnetic material with a point group that includes an $E$-type irrep ($E$, $E_g$, or $E_u$) consisting of two 1D irreps when complex representations are allowed, indicating that similar phonon energy splittings can be expected in such materials. 

The phonon angular momentum is calculated using \eqs{lz-atom}{lz-total}, and the results are listed in Table~\ref{tab:res-cri3-3d-e}. Due to the $C_3$ symmetry, the total angular momentum of a given chiral phonon can only have a $z$-component.  We find values of $|L_z|$ ranging from 0.13 to 0.95 in units of $\hbar$, indicative of a substantial angular momentum for many of the chiral phonons, especially the higher-frequency ones.

We find that the bare circular modes, constructed as $\ket{u_\pm}=(\ket{u_1}\mp i\ket{u_2})/\sqrt{2}$ from the real bare modes, can be used to evaluate the PAM to a good approximation. The results are close to those shown in Table~\ref{tab:res-cri3-3d-e}, with a typical error of $10^{-3} \hbar$ in $|L_z|$. However, when calculating the PAM in this way, the angular momentum vanishes exactly when summed over any pair of chiral phonons, which have equal and opposite $L_z$ values. In contrast, solutions using the full matrices of \eq{eom-full} reveal that the cancellation is no longer perfect, and the total angular momentum of a pair is small but non-zero. The slight discrepancy between the two $|L_z|$ values arises from the magnon-mediated phonon-phonon mixing, which appears at second-order in the phonon-magnon coupling and therefore makes only a minor contribution. 
We provide a more detailed discussion on this topic based on a perturbation approach in Appendix~\ref{ap:pert-ksp}. 

%=================================================
\subsubsection{Importance of $\gpp$ and $\gps$ in calculating energies}
\seclab{cri3-comp}
%=================================================

\begin{table}
\caption{\label{tab:res-cri3-3d-a} Computed properties of zone-center $A_g$ and $A_u$ phonons in bulk CrI$_3$. Energy shifts $\Delta E$ are defined relative to the bare phonon energies $E_0$. $L_z$ is the phonon angular momentum of \eq{lz-total}.}
\begin{ruledtabular}
\begin{tabular}{cddd}
Irrep & \multicolumn{1}{c}{$E_0$ (meV)} & \multicolumn{1}{c}{$\Delta E$ ($10^{-8}$~meV)} & \multicolumn{1}{c}{$L_z\,(10^{-4} \hbar)$}  
\\
\hline
$A_g$ & 9.95 & -8.8 & 1.99  
\\
& 11.35 &  2.6 & -2.37 
\\
& 16.50 & -50.6 & -1.07 
\\
& 26.51 & 98.6 & 1.96 
\\
%\hline
$A_u$  & 8.13 & -4.5 & -1.16 
\\
& 16.60 & -7.9 & 2.65
\\
& 31.74 & 32.7 & -0.52 
\end{tabular}    
\end{ruledtabular}
\end{table}

The methodology of Ref.~\cite{bonini2023} for CrI$_3$ differs from the approach of this work (i.e., \eq{eom-full} and \eq{full-mat}) in the following ways. First, the only Berry curvature considered in \eq{full-mat} was $\gss$. Second, the minimal model in Ref.~\cite{bonini2023} neglects the mixing between bare phonon doublets, considering the interaction which each doublet and one magnon branch separately. In Appendix~\ref{ap:sp-model}, we discuss a more comprehensive version of that model which still neglects $\gpp$, $\gsp$, and $\gps$, but incorporates all bare phonons and magnons. 
To show the importance of $\gpp$ and $\gsp$ ($\gps$) in accurately determining energy splittings, we also apply the model present in Appendix~\ref{ap:sp-model} to calculate chiral phonon energies for $E_g$ and $E_u$ modes, and the results are present in Table~\ref{tab:res-cri3-3d-e}. 

As shown in Table~\ref{tab:res-cri3-3d-e}, both methods exhibit an energy splitting of chiral modes compared to bare phonons. The energy splitting of low-energy $E_g$ and $E_u$ phonons is similar to that obtained using the present approach, suggesting that $\gpp$ and $\gps$ have little effect on those modes. However, for high-energy $E_g$ and $E_u$ phonons, the spin-phonon model results differ significantly, indicating the importance of $\gpp$ and $\gps$ for these phonons. 
Specifically, the absolute values of the $\gpp$ matrix elements for these four $E_g$ modes are \{0.995, 0.254, 0.972, 5.623\} in units of $\mu$eV. The highest-energy $E_g$ modes exhibit substantially greater values, leading to a pronounced energy splitting. Appendix~\ref{ap:pert-gpp} details the perturbative analysis of $\gpp$, with \eq{gpp-omega} specifying the energy corrections attributed to the $\gpp$ term. As indicated by \eq{gpp-omega}, $\gpp$ primarily accounts for the substantial splitting observed in the highest-energy $E_g$ mode, with both direct computation and perturbative treatment yielding a correction of $\pm0.0028$~meV. 

Interestingly, we find that the PAM calculated using the spin-phonon model (not shown) is in good agreement with the full method with an error of the order of $10^{-4}\,\hbar$, suggesting that the phonon vectors are not very different regardless of whether $\gpp$ and $\gps$ are included or not.

We now turn to a consideration of the $A_g$ and $A_u$ modes.
The spin-phonon model presented in Appendix~\ref{ap:sp-model} has no effect on the $A_g$ and $A_u$ phonons, since $\ksp$ gives the coupling to magnons, and there are no zone-center magnons in these symmetry sectors. Considering the present approach, although all (ss) and (sp) terms are absent in the $A_g$ and $A_u$ sectors, the $\gpp$ matrix does not vanish, and causes a slight mixing between two different $A_g$ (or $A_u$) bare phonons. That is, we can write $\ket{u_n}\simeq\ket{u_n^{(0)}}+i \sum_m \delta_m\ket{u_m^{(0)}}$, 
where $\ket{u_{\{n,m\}}^{(0)}}$ represents the bare phonons from $A_g$ (or $A_u$) irrep. Here, $i\delta_m$ is purely imaginary and expected to be small, attributed to the minor size of $\gpp$ matrix element, as discussed in Appendix~\ref{ap:pert-gpp}. 

This results in a nonzero PAM and nonzero energy shift of each phonon relative to its bare energy. Although these phonons exhibit non-zero PAM, we refrain from referring to them as ``chiral phonons" because they belong to the $\chi(C_3)=1$ sector of modes that are invariant under the $C_3$ operation. Instead, we refer to them as $(1+i\delta)$-type perturbed phonons, or simply perturbed phonons. 

The energy shifts and PAM for $A_g$ and $A_u$ perturbed phonons are presented in Table~\ref{tab:res-cri3-3d-a}, and the real and imaginary parts of the $A_g$ perturbed phonon near $10$~meV ($A_g^{(1)}$) are visualized in \fref{vis-ag}.

\begin{figure}
\centering\includegraphics[width=\columnwidth]{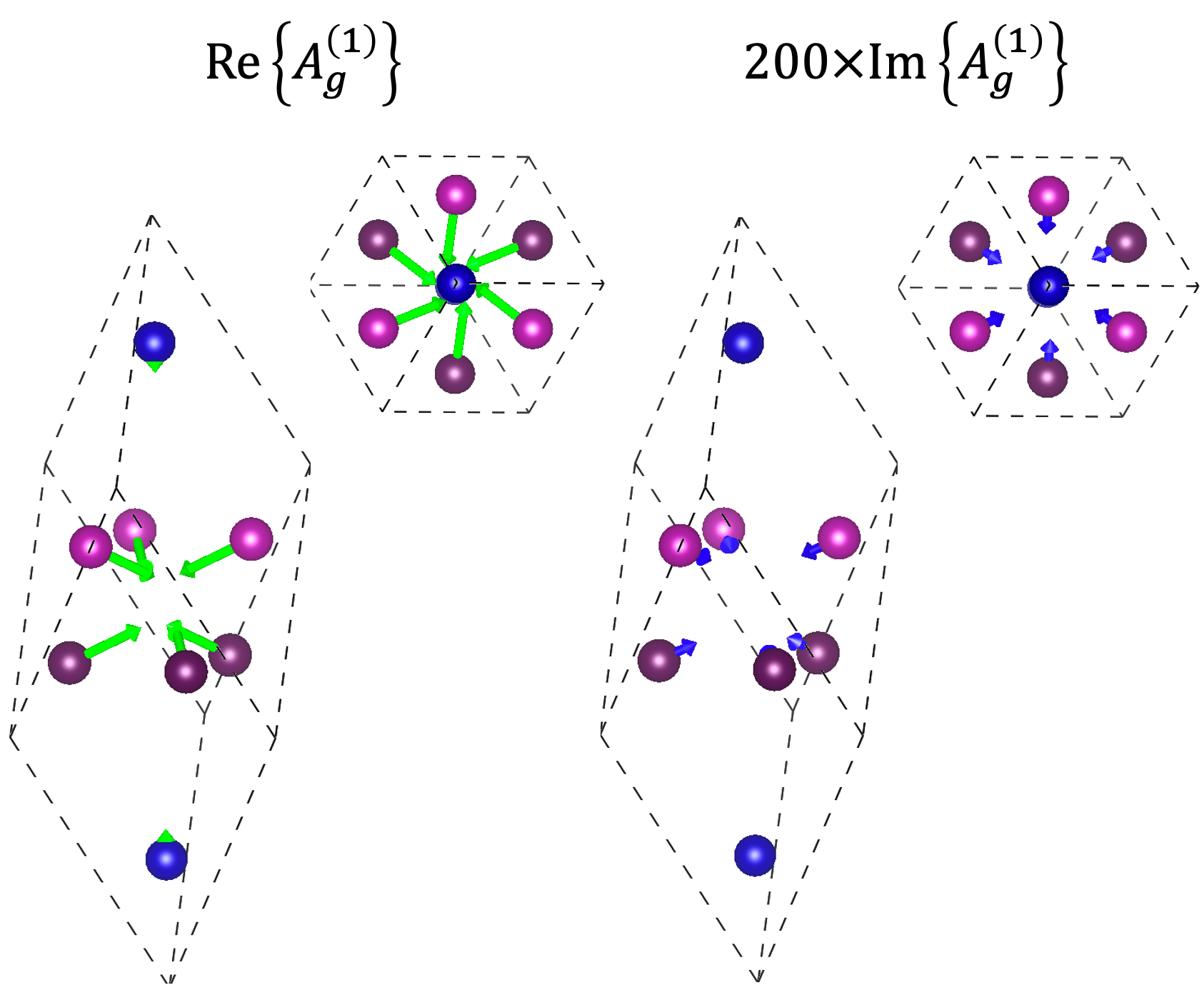}
\caption{Visualization of the real and imaginary components of the perturbed phonon $A_g^{(1)}$. Cr atoms are depicted in blue, while top-layer and bottom-layer I atoms are in bright and dark magenta, respectively. The green vectors represent the real part of the phonon displacement, and the blue vectors represent the imaginary part, which are amplified 200 times.}
\figlab{vis-ag}
\end{figure}

%=================================================
\subsubsection{Bare and perturbed magnons}
\seclab{cri3-magnon}
%=================================================

So far we focused solely on the phonon-like solutions of the equation of motion given by \eq{eom-full} in the present approach (or \eq{eom-sp} for the spin-phonon model). Nevertheless, these equations also admit magnon-like solutions, which we call ``perturbed magnons.'' In this section, we investigate the impact of phonons on the magnon spectrum, using bulk CrI$_3$ as a case study. The numerical values for the matrices $\kss$ and $\gss$ are provided in Appendix~\ref{ap:kss}. 

The EOM for a bare magnon without any phonon-magnon interaction is given by \eq{eom-magnon}. The energies of bare magnons are listed in Table~\ref{tab:cri3-magnon}. The $E_g$ magnon corresponds to the acoustic mode, where the two Cr spins have the same canting, while the $E_u$ magnon corresponds to the optical mode, where the two Cr spins have opposite canting. Both magnons have positive energies and belong to the `$+$' sector with $\chi(C_3)=\varepsilon$, indicating that the two Cr spins are rotating clockwise.%
\footnote{It is important to clarify that the reference to the clockwise rotation of the two spins is in the context of their motion as viewed from the direction of $s_z$. However, viewed from the direction of $M_z$, which is opposite to $s_z$, the two magnetic moments still rotate counterclockwise in the ($M_x$, $M_y$) plane.}

The perturbed magnons have slightly modified energies (Table~\ref{tab:cri3-magnon}). The energy change from the bare magnon for the $E_u$ mode is more significant compared to the $E_g$ magnon, since the $E_u$ phonons are closer to the $E_u$ magnon in the energy spectrum. 
By far the most significant contribution to the energy shifts is the inclusion of $\ksp$; neglecting the $\gpp$, $\gsp$, and $\gps$ terms changes the energies by less than 0.4 $\mu$eV. 
A perturbation treatment of magnon energies is provided in Appendix~\ref{ap:pert-ksp}. 

\begin{table}
\caption{\label{tab:cri3-magnon} Energies of the bare magnons $E_0$, energy shifts of perturbed magnon modes comparing to bare magnons $\Delta E$, and the corresponding $C_3$ eigenvalues $\chi(C_3)$ for both $E_g$ and $E_u$ modes in bulk CrI$_3$.}
\begin{ruledtabular}
\begin{tabular}{cccc}
Irrep & $E_0$~(meV)  & $\Delta E$~(meV)  & $\chi(C_3)$   \\
\colrule
$E_g$ & $\phm 0.5902$  & $-0.0046$ & $\varepsilon$ 
\\
$E_u$ & $22.8635$ & $-0.0224$ & $\varepsilon$
\end{tabular}    
\end{ruledtabular}
\end{table}

Note that the equations of motion have two additional negative-energy magnon solutions belonging to the `$-$' sector with $\chi(C_3)=\varepsilon^*$. These correspond to the counterclockwise precession of spins and are not physically observable. Nonetheless, they still influence the system dynamics through their interaction with physical `$-$' phonons, which acquire some magnon dressing in which the spin vectors are forced to precess in the unnatural counterclockwise sense. This becomes clear from the perturbation analysis presented in Appendix~\ref{ap:pert}, where the summation index $\mu$ in \eq{un-1st} and \eqr{ph-en-xi}{ph-en-2nd} runs over all solutions of \eq{smu-def}, including those of negative energy. However, since the energy denominators in \eq{ph-en-2nd} are larger when coupling to negative-energy solutions, the magnon dressing is typically smaller. This explains why the `$+$' phonons, which couple to positive-energy magnon solutions, are more strongly perturbed than the `$-$' ones, which do not. This can be seen in Table~\ref{tab:res-cri3-3d-e}, where it is especially noticeable for the $E_u$ modes.

%=================================================
\subsection{Phonons in antiferromagnetic Cr$_2$O$_3$}
\seclab{afm-3d}
%=================================================

In this section, we present our results for the phonons in bulk Cr$_2$O$_3$. As shown in \fref{cr2o3-crystal}, Cr$_2$O$_3$ has an AFM insulating ground state with antiparallel magnetic moments aligned along the threefold-rotational $z$ axis.

\begin{figure}
\centering\includegraphics[width=0.6\columnwidth]{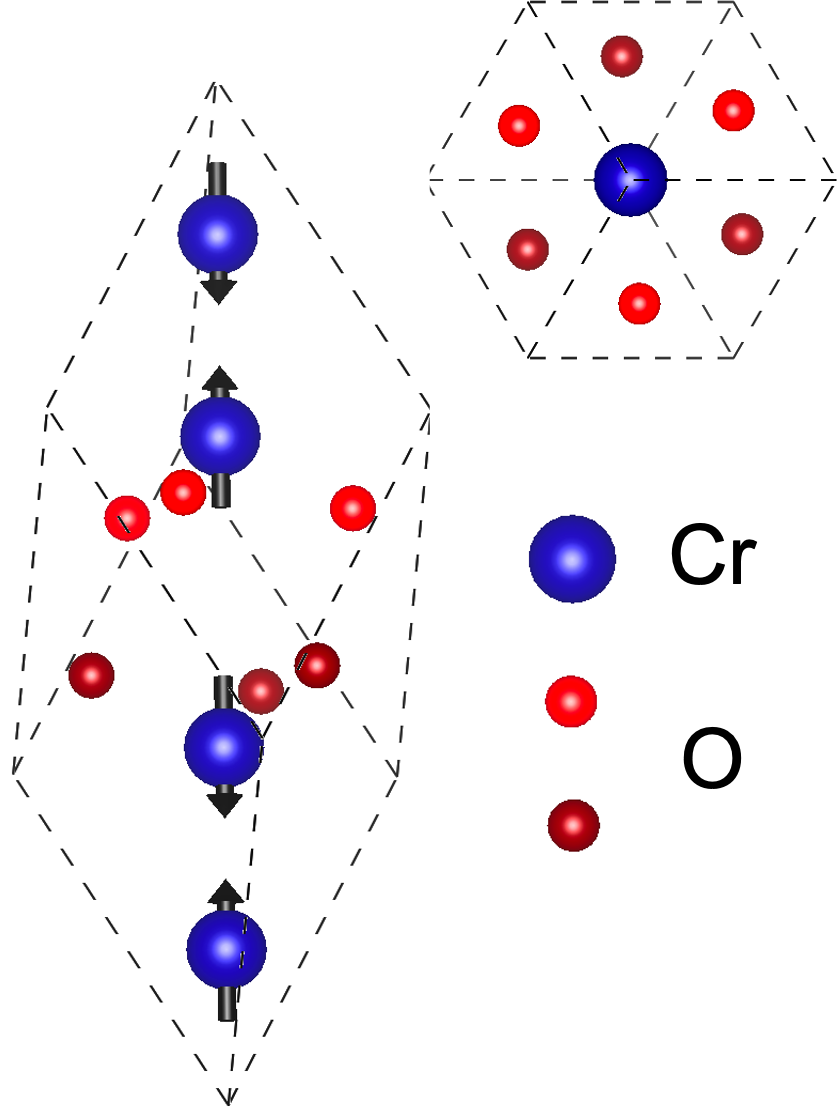}
\caption{A visualization of the crystal structure and magnetic moments of the unit cell of bulk Cr$_2$O$_3$. The Cr atoms are depicted in blue, while the top-layer and bottom-layer O atoms are shown in bright and dark red, respectively. The black vectors indicate the magnetic moments, which are oriented along the $z$-direction.}
\figlab{cr2o3-crystal}
\end{figure}

The magnetic symmetry of Cr$_2$O$_3$ is richer than that of CrI$_3$, in that it belongs to a Type III (``black-white'') magnetic group~\cite{dresselhaus2007}.  Since we are interested in zone-center phonons, we frame our discussion in terms of the magnetic point group $\mathcal{G}$. The black-white character means that half the the elements of $\mathcal{G}$ come with the TR operator $\mathcal{T}$ and half come without. The latter form a ``unitary subgroup'' $\mathcal{H}$, and the others are of the form $g\mathcal{H}$ where $g$ is one of the antiunitary operators in $\mathcal{G}$. The ``structural point group'' $\bar{\mathcal{G}}$ contains the same list of operators as in $\mathcal{G}$, except that $\cal T$ is removed from all the antiunitary operators. The important point in what follows is that while the Hessian matrices are even under all elements of $\bar{\mathcal{G}}$, the Berry curvature matrices are even under the elements of $\mathcal{H}$ and odd under the elements of $\bar{\mathcal{G}}-\mathcal{H}$. As a consequence, when the Berry curvature is included, the irreps of unitary subgroup $\mathcal{H}$, not the structural group $\bar{\mathcal{G}}$, should be used to label the perturbed modes of this system.

In the case of Cr$_2$O$_3$, the magnetic space group is $R\bar{3}'c'$, and the structural point group is $D_{3d}$. The magnetic ordering on the Cr sublattice breaks inversion ($i$), dihedral mirror ($\sigma_d$), and rotoinverion ($S_6$) symmetries, reducing the unitary point group to $D_3$, whose irreps will be used to label our perturbed phonons and magnons. The corresponding time-reversed operators $i\mathcal{T}$, $\sigma_d \mathcal{T}$, and $S_6 \mathcal{T}$ are present in the magnetic point group, but being antiunitary, do not induce any additional irreps. 

While the experimental magnetic moment of Cr$_2$O$_3$ is found to be along the $z$ direction, DFT calculations predict an in-plane magnetic moment. This discrepancy can be resolved by applying a 2\% epitaxial strain, which results in an easy-axis ground state~\cite{mu2019}. To ensure consistency with the experimental ground state, we apply a 2\% epitaxial strain in our calculations, as our model assumes spins to be oriented along the $z$-direction. 

As previously discussed, the irreps of point group $D_3$ should be used to characterize zone-center perturbed phonons, which can be decomposed into $4 A_1\oplus 6 A_2 \oplus 10 E$. Among them, one $A_2$ and two $E$ modes correspond to acoustic phonons. According to the character table of $D_3$, the $E$ irreps have to be 2D irreps, indicating that there is no degeneracy breaking or energy splitting compared to bare phonons. This is confirmed by the numerical results of perturbed phonon energies, which are presented in Table~\ref{tab:res-cr2o3}. It is worth noting that the perturbed phonon energies differ from those of the bare phonons, which are also included in Table~\ref{tab:res-cr2o3} for reference purposes. 

\begin{table}
\caption{\label{tab:res-cr2o3} Bare phonon irreps, energies, atom-resolved phonon angular momentum (ARPAM) of Cr atoms in $z$-direction (Cr $L_z$), ARPAM of O atoms in $x$-direction (O $L_x$), and inter-irrep mixing ($\rho$) of perturbed phonons in bulk Cr$_2$O$_3$. The bare phonon energies $E_0$ are included as a reference for the energy shifts $\Delta E$. $E$-irrep ($E_g$ and $E_u$) phonons are doubly degenerate. Note that `$0$' denotes an entry that is precisely zero by symmetry, while `$0.000$' signifies a non-zero entry that has been rounded to zero. The energy shifts $\Delta E$ for $A$-irrep perturbed phonons are on the order of $10^{-4}\;\mu\mbox{eV}$, and $L_x$ for the O atom of the $A_{2g}^{(1)}$ phonon at $33.310$~meV is $9 \times 10^{-9}\hbar$.}
\begin{ruledtabular}
\begin{tabular}{cddddd}
    Irrep & \multicolumn{1}{c}{$E_0$} & \multicolumn{1}{c}{$\Delta E$} & \multicolumn{1}{c}{\textrm{Cr} $L_z$} & \multicolumn{1}{c}{\textrm{O} $L_x$} & \multicolumn{1}{c}{$\rho$} \\
    & \multicolumn{1}{c}{\textrm{(meV)}} & \multicolumn{1}{c}{($\mu$\textrm{eV})} & \multicolumn{1}{c}{$(10^{-4}\hbar)$} &  \multicolumn{1}{c}{$(10^{-4}\hbar)$} & \multicolumn{1}{c}{$(10^{-4})$} \\
\colrule
$E_g$   &  36.458 &  -0.399 &  -2.087 &  -0.227 &   4.319  \\
        &  43.739 & -1.738 &   1.302 &   0.171 &   2.219 \\
        &  49.602 & -0.438 &  -0.073 &   0.175 &   1.104 \\
        &  65.043 & -0.124 &  -0.125 &  -0.037 &   2.592 \\
        &  76.667 &  0.355 &   0.583 &  -1.471 &  10.547 \\
$E_u$   &  38.209 &  -0.010 &   0.485 &   0.213 &   4.877 \\
        &  55.803 &  -0.095 &   0.231 &   0.314 &   1.086 \\
        &  67.431 &   1.161 &   1.537 &  -0.012 &   5.785 \\
        &  76.041 &   0.023 &  -0.064 &   0.371 &  10.231 \\
\colrule
$A_{1g}$ &  36.326 &   0.000 &   0     &   0     &   0.225 \\
         &  67.779 &   0.000 &   0     &   0     &   0.219 \\
%\colrule
$A_{1u}$ &  51.723 &   0.000 &   0     &   0     &   0.335 \\
         &  77.221 &   0.000 &   0     &   0     &   0.203 \\
\colrule
$A_{2g}$ &  33.310 &   0.000 &   0     &   0.000 &   0.727 \\
         &  56.989 &   0.000 &   0     &   0.517 &   2.270 \\
         &  83.594 &   0.000 &   0     &  -0.161 &   1.319 \\
%\colrule
$A_{2u}$ &  50.565 &   0.000 &   0     &   0.090 &   1.784 \\
         &  67.760 &   0.000 &   0     &  -0.604 &   2.212 \\
\end{tabular}    
\end{ruledtabular}
\end{table}

%=================================================
\subsubsection{Mixing of phonon irreps} 
%=================================================
 
As mentioned earlier, although $i$, $\sigma_d$, and $S_6$ are no longer symmetries, $i\mathcal{T}$, $\sigma_d \mathcal{T}$, and $S_6\mathcal{T}$ remain symmetries of the system. On the other hand, the matrices $\kpp$ and $\Mp$ are both real. Applying the symmetry operations $i\mathcal{T}$, $\sigma_d \mathcal{T}$, and $S_6\mathcal{T}$ to these matrices is equivalent to applying $i$, $\sigma_d $, and $S_6$, respectively. Therefore, $\kpp$ and $\Mp$ are symmetric not only under the $D_3$ symmetry operations, but also under $i$, $\sigma_d$, and $S_6$, resulting in a point group symmetry of $D_{3d}$.

This has an important impact on the symmetries of zone-center bare phonons, which are calculated using $\kpp$ and $\Mp$: bare phonons at the $\Gamma$ point must have $D_{3d}$ symmetry, and thus can be labeled by the irreps of $D_{3d}$. The $\Gamma$ point bare phonons decompose into $5 E_g \oplus 5 E_u \oplus 2A_{1g} \oplus 2 A_{1u} \oplus 3 A_{2g} \oplus 3 A_{2u}$, where one $E_u$ doublet and one $A_{2u}$ mode are acoustic modes. The irreps of the bare phonons are included in Table~\ref{tab:res-cr2o3}. 

The presence of magnetic moments in bulk Cr$_2$O$_3$ results in the breaking of $i$, $\sigma_d$, and $S_6$ symmetries associated with the bare phonons. Consequently, the original $D_{3d}$ point group symmetry is reduced to $D_3$, leading to inter-irrep mixing between bare phonons of different irreps. Referring to the correlation table of the $D_{3d}$ point group, we find that the $E_g$ and $E_u$ irreps combine to form the $E$ irrep, the $A_{1g}$ and $A_{1u}$ combine to form the $A_1$ irrep, and the $A_{2g}$ and $A_{2u}$ irreps combine to form the $A_2$ irrep of $D_3$. An important question arises: How significant is this inter-irrep mixing, and to what extent does the spatial symmetry breaking due to the magnetic order impact the behavior of the phonons? 

To address this question, we project the bare phonons belonging to irreps of the $D_{3d}$ group onto the perturbed phonons. This allows us to determine the components of each irrep in $D_{3d}$ that contribute to the given perturbed phonon. We introduce the concept of irrep decomposition component, denoted as $\rho_n({\rm irrep})$, which is defined as
\begin{equation}
\rho_n({\rm irrep}) = \left(\sum_{m \in \mathrm{irrep}} \abs{\me{u_m^{(0)}}{\Mp}{u_n}}^2\right)^\frac{1}{2},
\eqlab{rho-irrep}
\end{equation}
where $n$ is the label for the perturbed phonon mode and $m$ runs over bare phonons from a particular irrep. 
The last column of Table~\ref{tab:res-cr2o3} reports our results for $\rho_n(E_u)$ for $n\in E_g$ and vice versa, and similarly for $A_{1g}$-$A_{1u}$ and $A_{2g}$-$A_{2u}$ components. In each case, the majority component (e.g., $\rho_n(E_g)$ for $n\in E_g$) is almost unity, so we list only the minority components.  These are non-zero as expected, but we find them to be quite small, typically between $10^{-3}$ and $10^{-4}$. This implies that the phonon sector exhibits only weak breaking of inversion symmetry, allowing us to continue to refer to the perturbed phonons as $E_g$-like or $E_u$-like.

%=================================================
\subsubsection{Experimental implications: Raman and infrared activity}
%=================================================

The experimental implication of the inter-irrep mixing is that the perturbed phonons will exhibit distinct Raman and IR activities compared to bare phonons, as summarized in Table~\ref{tab:cr2o3-irrep}. Above the N\'eel temperature, where TRS is preserved, the symmetry of the phonons corresponds to that of the bare phonons, so that $E_u$ phonons are IR-active but Raman-inactive. However, upon cooling the sample below the N\'eel temperature, $E_u$ phonons undergo mixing with $E_g$ phonons, resulting in the emergence of $E_u$-like perturbed phonons. These $E_u$-like phonons possess Raman activity (in addition to IR activity), as they technically belong to the $E$ irrep of the $D_3$ point group. However, since the inter-irrep mixing is small, the Raman activity of the $E_u$-like phonons is relatively weak. Nonetheless, recent Raman measurements have confirmed the presence of these features~\cite{wu2023}. In a similar way, the Raman-active $E_g$-like phonons acquire some small IR activity. 
Clearly, an approach such as ours, which treats the coupling of phonons and spins in a realistic and symmetry-consistent manner, is needed to describe these effects.

\begin{table}
\caption{\label{tab:cr2o3-irrep} Raman and infrared (IR) activities for bare and perturbed phonons. For perturbed phonons, the Irrep$^*$ column indicates the irrep of perturbed phonons labeled by irreps of $D_{3d}$, which is made possible by the weak inter-irrep mixing. In cases where a perturbed phonon exhibits both Raman and IR activity, the minor activity is presented within parentheses and is subordinate to the major activity.}
\begin{ruledtabular}
\begin{tabular}{ccccc}
\multicolumn{2}{c}{Bare phonons}  &  \multicolumn{3}{c}{Perturbed phonons} \\
Irrep & Activity & Irrep & Irrep$^*$ & Activity \\
\colrule
$E_g$ & Raman & $E$ & $E_g$-like & Raman (IR) \\
$E_u$ & IR    & $E$                  & $E_u$-like & IR (Raman) \\
%\colrule
$A_{1g}$ & Raman & $A_1$ & $A_{1g}$-like & Raman \\
$A_{1u}$ &       & $A_1$ & $A_{1u}$-like & Raman \\
%\colrule
$A_{2g}$ & & $A_2$ & $A_{2g}$-like & IR \\
$A_{2u}$ & IR  & $A_2$  & $A_{2u}$-like & IR \\
\end{tabular}    
\end{ruledtabular}
\end{table}

%=================================================
\subsubsection{Atom-resolved phonon angular momentum}
%=================================================

As mentioned earlier, inversion is no longer a symmetry in Cr$_2$O$_3$, but inversion times time reversal ($i\mathcal{T}$) remains a symmetry. This means that $i\mathcal{T}$ maps total angular momentum from $\vec{L}$ to $-\vec{L}$, resulting in $\vec{L} = \vec{0}$ for any non-degenerate single mode or for the sum over two degenerate modes. However, each atom can still possess a nonzero atom-resolved phonon angular momentum (ARPAM), defined by \eqs{lz-atom}{lz-total} above. ARPAM is a pseudovector assigned to each atom and has the same symmetry as a local magnetic moment. The configuration of the local magnetic moment is determined by the magnetic space group, and this is also true for the ARPAM. 

For Cr$_2$O$_3$, Cr and O atoms occupy Wyckoff positions $12c$ and $18e$ respectively. The possible configurations of ARPAM (and local magnetic moments) are listed in Table~\ref{tab:cr2o3-config} \footnote{Based on the \textit{Bilbao crystallography server}~\cite{aroyo2006-1,aroyo2006-2,aroyo2011} (through the MWYCKPOS module~\cite{gallego2012}).}, indicating that Cr atoms can only have out-of-plane angular momentum $L_z$ or $-L_z$, while O atoms can only have in-plane angular momentum in a way that respects $C_3$ symmetry. This is further supported by the numerical results presented in Table~\ref{tab:res-cr2o3}. In \fref{cr2o3-arpam}, we visualize the ARPAM for the $E_g^{(1)}$ perturbed phonon around 36~meV and the $A_{2g}^{(2)}$ perturbed phonon near 57~meV, where the $L_z$ for Cr atoms and the in-plane angular momentum for O atoms are clearly visible. 

\begin{table}
\caption{\label{tab:cr2o3-config} Possible configurations of atom-resolved phonon angular momentum (ARPAM) for bulk Cr$_2$O$_3$ according to the magnetic space group $R\bar{3}'c'$. Cr atoms can only have out-of-plane angular momentum $L_z$ or $-L_z$, while O atoms can only have in-plane angular momentum and must respect $C_3$ symmetry.}
\begin{ruledtabular}
\begin{tabular}{ccc}
Atom & Wyckoff positions & ARPAM \\
\colrule
Cr &  $12c$ & $(0,0,L_z)$, $(0,0,-L_z)$ \\
O  &  $18e$ & $\{I,C_3,C_3^2 \} \cdot (L_x, 0, 0)$
\end{tabular}    
\end{ruledtabular}
\end{table}

It is important to note that there exists a gauge freedom for each $E$ doublet, where the two degenerate modes can be unitarily mixed with each other through a $U(2)$ matrix. Therefore, discussing ARPAM for each individual $E$ mode is meaningless, as it is gauge-dependent. However, the sum of ARPAM for the two degenerate $E$ modes is gauge-independent. To demonstrate this, let us consider an $E$ doublet labeled by $n=\{1,2\}$, where $L_{n,Iz}$ denotes the ARPAM for atom $I$ along the $z$-direction. Then, the sum of $L_{1,Iz}$ and $L_{2,Iz}$ is
\begin{align}
    & \; L_{1, I z} + L_{2, I z} \nn
    & = \hbar M_I \Tr
    \Bigr[  \begin{pmatrix}
        u_{1, I x}^* & u_{1, I y}^* \\
        u_{2, I x}^* & u_{2, I y}^* 
    \end{pmatrix}
    \begin{pmatrix}
        0 & -i \\
        i & 0
    \end{pmatrix}
    \begin{pmatrix}
        u_{1, I x} & u_{2, I x}\\
        u_{1, I y} & u_{2, I y}
    \end{pmatrix} \Bigr] \nn
    & = 2 \hbar M_I [\Im(u_{1, I x}^* u_{1, I y}) + \Im(u_{2, I x}^* u_{2, I y}) ]\,,
    \eqlab{lz-sum}
\end{align}
which is the trace of a product of three matrices. While different gauges correspond to different bases for these matrices, they do not affect the trace, making $L_{1,Iz} + L_{2,Iz}$ a gauge-invariant quantity. This is also true for the $x$ and $y$ directions. The ARPAM for $E$-irrep perturbed phonons shown in Table~\ref{tab:res-cr2o3} is traced over the two degenerate modes in the doublet. 
For $E$ modes with nonzero ARPAM, there is no gauge choice in the degenerate subspace where the phonon mode displacements can be expressed as purely real eigenvectors. 

\begin{figure}
\centering\includegraphics[width=\columnwidth]{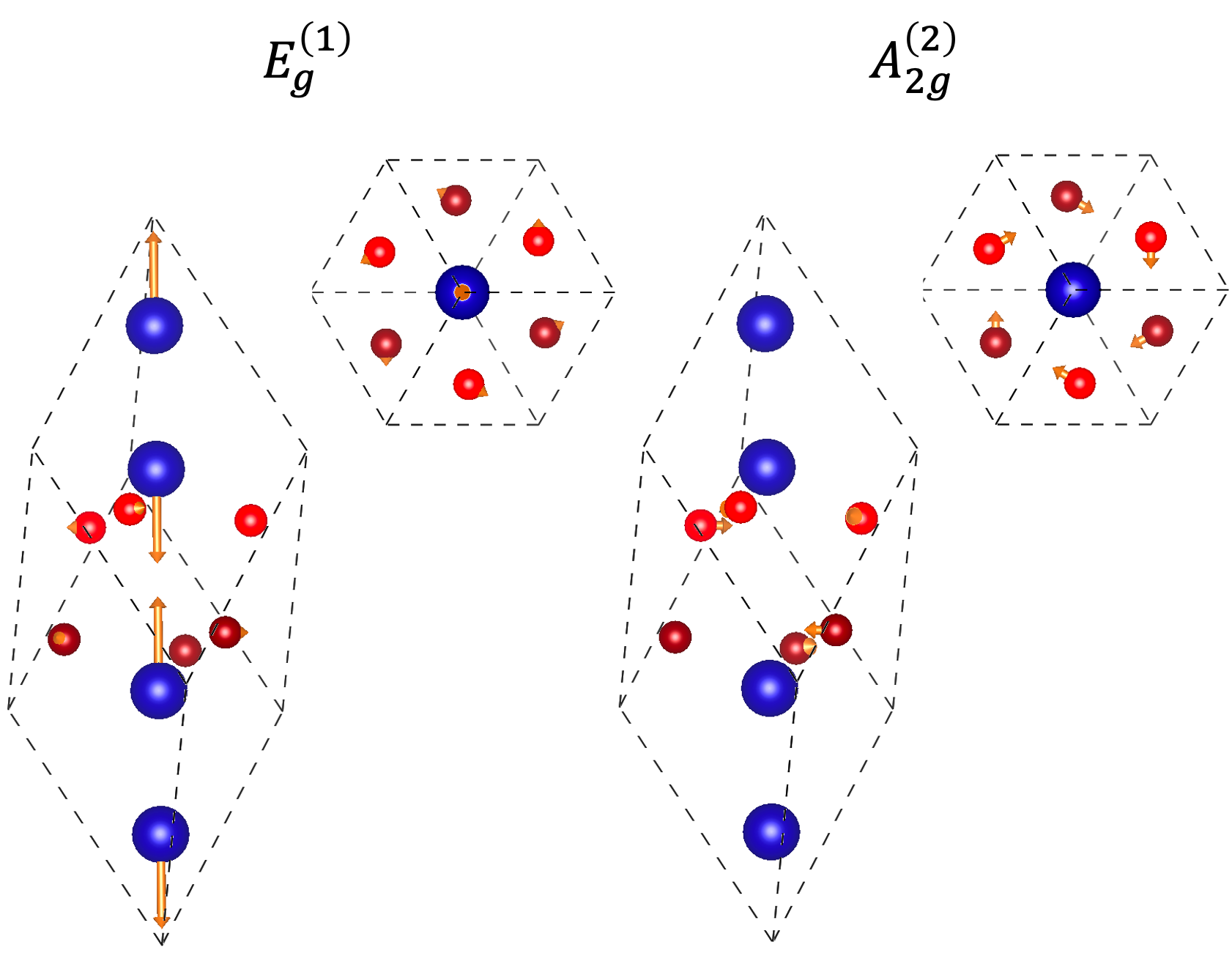}
\caption{Atom-resolved phonon angular momentum (ARPAM) for two selected modes in bulk Cr$_2$O$_3$: $E_g^{(1)}$-like mode near 36~meV and $A_{2g}^{(2)}$-like mode near 57~meV. The ARPAM for the $E_g$-like mode is traced over two degenerate modes. The amplitude of ARPAM in this figure is amplified by a factor of 1000 compared to \fref{cr2o3-eg}.}
\figlab{cr2o3-arpam}
\end{figure}

%=================================================
\subsubsection{Chiral decomposition of $E$ modes}
%=================================================

Despite the fact that perturbed $E$ phonons are always doubly degenerate, it is still worthwhile to decompose the doublet into two single $E$ chiral phonons that respect $C_3$ symmetry individually and have complex eigenvalues. There are several reasons for this. Firstly, these chiral phonons are excited by circular polarized photons~\cite{zhang2015}. Secondly, they exhibit different energies in the presence of an external magnetic field along the $z$-direction. Finally, these modes correspond to a special gauge choice in which each individual mode has the largest $L_z$ magnitude. To decompose the $E$ doublets, we diagonalize the $C_3$ matrix using the bases formed by two degenerate $E$ modes. The resulting eigenvectors correspond to two $E$ chiral phonons that respect $C_3$ symmetry individually. Since the two $E$ modes together respect the $C_3$ symmetry, the $C_3$ matrix has to be a $2 \times 2$ unitary matrix, and is therefore diagonalizable. 

We have decomposed the $E_g$-like perturbed phonons around $36$~meV into two chiral phonons, denoted as $E_g^{(1)+}$ and $E_g^{(1)-}$, and have visualized their real parts, imaginary parts, and ARPAM in \fref{cr2o3-eg}. The real parts of the two modes are almost identical to each other, while the imaginary parts are almost opposite. As a consequence, the ARPAMs of the two modes are also nearly opposite to each other. The PAMs for the `$+$' and `$-$' modes are $\pm 0.8593\hbar$, respectively. However, the ARPAMs of the two modes do not cancel out perfectly, and their sum is shown in the left panel of \fref{cr2o3-arpam}. Other $E_g$-like and $E_u$-like perturbed phonons can also be decomposed in the same manner. 

\begin{figure}
\centering\includegraphics[width=\columnwidth]{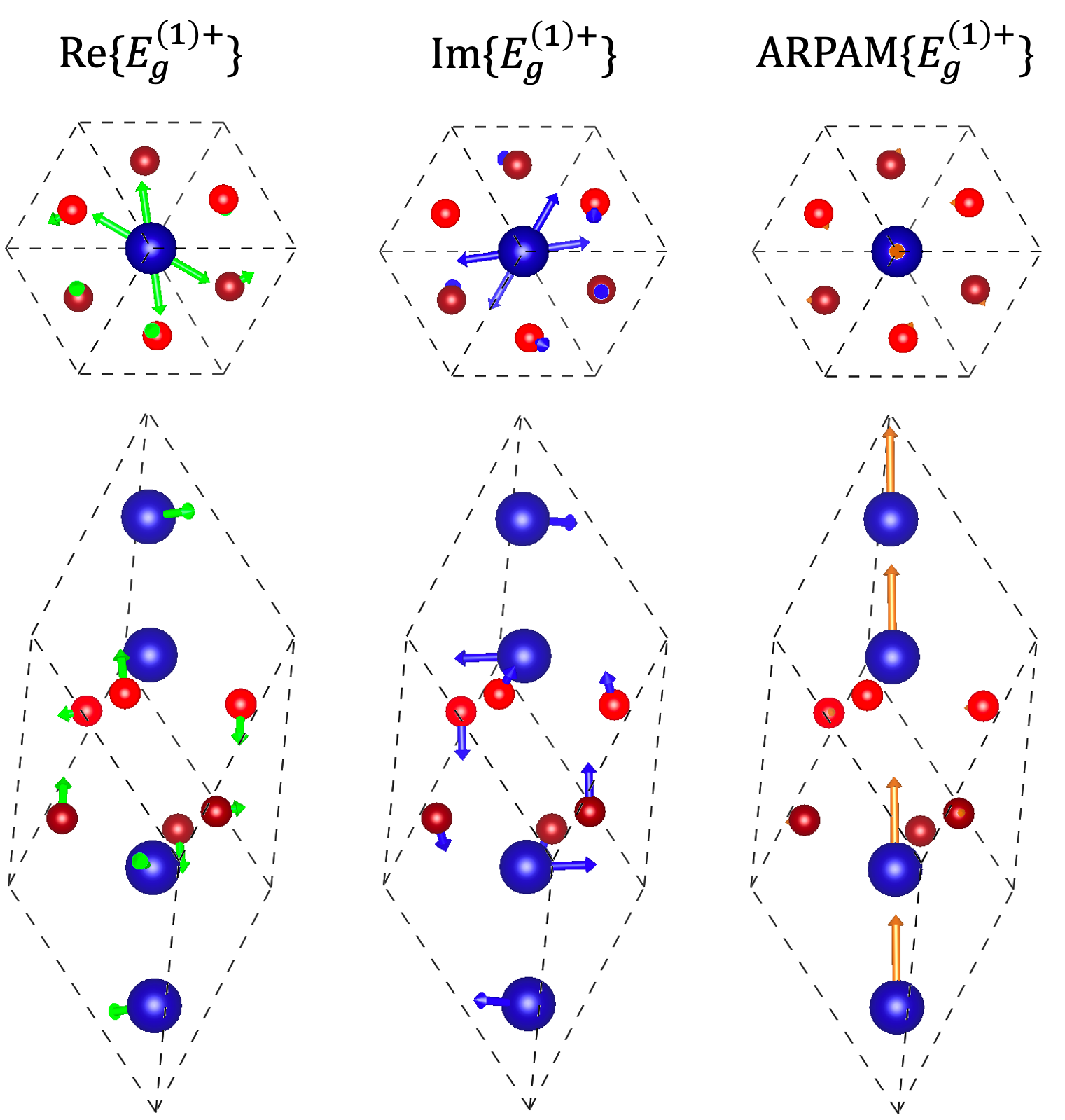}
\caption{The real and imaginary components, as well as atom-resolved phonon angular momentum (ARPAM), of the chiral phonon $E_g^{(1)+}$ around 36~meV in Cr$_2$O$_3$. The Cr atoms are shown in blue, while the top-layer and bottom-layer O atoms are depicted in bright and dark red, respectively. The green vectors indicate the real part of the phonon displacement, the blue vectors indicate the imaginary part, and the brown vectors represent the ARPAM. The real parts of $E_g^{(1)-}$ are almost identical to $E_g^{(1)+}$, while its imaginary parts and ARPAM are almost opposite to those of $E_g^{(1)+}$. }
\figlab{cr2o3-eg}
\end{figure}

%=================================================
\subsubsection{Magnons}
%=================================================

We now turn to a discussion of the magnons. As a reminder, bare magnons are solved as described in \sref{cri3-magnon}, while perturbed magnons are magnon-like solutions of the EOM of the present approach. 
The energies of bare and perturbed magnons are presented in Table~\ref{tab:cr2o3-magnon}, and the numerical values for the matrices $\kss$ and $\gss$ describing the bare magnons are given in Appendix~\ref{ap:kss}.

We find that bare and perturbed magnons have almost identical energies, which results from the weak SOC for both the Cr and the O atoms.  
Since magnons belong to the $E$ irrep, they must be doubly degenerate. These degenerate magnons can be decomposed into two magnons, each with $\chi(C_3)$ equal to $\varepsilon$ or $\varepsilon^*$, following the same method we used with $E$ phonons. Additionally, these magnons can also be excited by circular polarized photons. 

We next analyze the symmetries of magnons
by considering the $E_g$ and $E_u$ components of each magnon mode, in a manner analogous to \eq{rho-irrep}. Unlike the bare phonons, however, the bare magnons do not have well-defined parity, because the magnetic order strongly violates inversion symmetry. Therefore, we base our analysis on the eigenvectors of the anisotropy matrix $K^{(ss)}$ instead.  Note that $\kss$, being quadratic in the spin DOF, is real and symmetric, so that the symmetry operator $i\mathcal{T}$ behaves like $i$ for $\kss$. Thus, it has well defined $E_g$ and $E_u$ eigenvectors that we denote as $\ket{t_\mu}$. 
Then the $E_g$ and $E_u$ components of a general solution $\ket{s_\mu}$ are defined via
\begin{equation}
    \rho_\mu({\rm{irrep}}) = \left(\sum_{\nu \in \mathrm{irrep}} |\ip{t_\nu}{s_\mu}|^2  \right)^{1/2} \,,
\end{equation}
where both $\ket{t_\nu}$ and $\ket{s_\mu}$ are normalized.
These $\rho$ values can be used to quantify the extent to which the inversion symmetry is broken for magnons, and the numerical values are listed in Table~\ref{tab:cr2o3-magnon}. We find that the acoustic magnons, which have lower energies, are nearly $E_u$ modes. However, the optical magnons have comparable $\rho(E_g)$ and $\rho(E_u)$ values, indicating that the inversion symmetry is strongly broken for these modes. This is not surprising, and is true also for the bare magnons, since the equations of motion for spin strongly violate TRS.

Note that we apply a 2\% epitaxial strain to fix the sign of the magnetic anisotropic energy, which is related to the energy of acoustic magnons. Although the variation of the magnon energies is sensitive to strain, the impact of such variations on the phonon energy corrections is negligible. This is clarified by the perturbative analysis in Appendix~\ref{ap:pert}, particularly evident in \eq{ph-en-2nd}. Given that the energy difference between the $E$ phonons and acoustic magnons is over $30$~meV, a $1$~meV variation in the magnon energy leads to a change of only about 3\% in the phonon energy correction. Thus, changes in acoustic magnon energy, or the magnetic anisotropic energy, do not significantly influence the phonon energy corrections. 

\begin{table}
\caption{\label{tab:cr2o3-magnon} Energies of the bare magnons ($E_0$) and perturbed magnons ($E$) in bulk Cr$_2$O$_3$, as well as their $E_g$ and $E_u$  components defined in \eq{rho-irrep}.}
\begin{ruledtabular}
\begin{tabular}{dddd}
\multicolumn{1}{r}{$E_0$~(meV)}  & \multicolumn{1}{r}{$\Delta E$~($\mu$eV)}  &  \multicolumn{1}{r}{$\rho(E_g)$}  &  \multicolumn{1}{r}{$\rho(E_u)$}   \\
\colrule
2.423   & -0.043  &  0.0386  &  0.9993  \\
66.899  & -0.045  &  0.6701  &  0.7423 
\end{tabular}    
\end{ruledtabular}
\end{table}

%=================================================
\subsubsection{Summary of Cr$_2$O$_3$}
%=================================================

In summary, our analysis shows that perturbed phonons possess $D_3$ symmetry, while bare phonons exhibit $D_{3d}$ symmetry. The $E$-type irreps of both $D_3$ and $D_{3d}$ are two-dimensional, implying that there is no energy splitting despite the breaking of TRS. However, an $E$ doublet can still be decomposed into two chiral phonons with different chiralities that respect $C_3$ symmetry. These chiral phonons can be excited by circular-polarized photons with different handedness, and their degeneracy will be lifted if an external magnetic field is present. Each perturbed phonon consists of components from two irreps of bare phonons, yet the level of inter-irrep mixing remains minimal. This characteristic implies that the inversion symmetry within the phonon sector is not strongly broken. Despite this minimal mixing, it is necessary to label perturbed phonons according to the irreps of the unitary subgroup of the magnetic point group. As a result, the infrared and Raman activity properties of these perturbed phonons differ from those of the bare phonons -- a phenomenon that is confirmed by recent experimental observations. 
Magnons belong to the $E$ irrep, and are therefore doubly degenerate. 
The magnons are mixtures of $E_g$ and $E_u$ sectors, indicating a strong inversion symmetry breaking for magnons. 

%=================================================
\subsection{Perturbed phonons and magnons in 2D systems}
\seclab{res-2d}
%=================================================

In \srefs{res-cri3}{afm-3d}, we have investigated chiral phonons in 3D systems, considering both FM and AFM cases. In this section, we shift our focus to 2D systems, specifically a monolayer of FM CrI$_3$ and a monolayer of AFM VPSe$_3$. 

%=================================================
\subsubsection{Chiral phonons in the monolayer CrI$_3$}
\seclab{cri3-2d}
%=================================================

Bulk CrI$_3$ has van der Waals gaps between layers, which allows it to be exfoliated to a 2D single layer while maintaining FM order~\cite{huang2017}. In this section, we report our calculations of chiral phonons in monolayer CrI$_3$, with a focus on comparing the results with those in the bulk cases. 

The crystal structure of monolayer CrI$_3$ is depicted in \fref{cri3-2d-crystal}. The magnetic group $P\bar{3}1m'$ is a Type-III black-white group, as was the case for bulk Cr$_2$O$_3$. The CrI$_3$ monolayer has higher structural symmetry than that of bulk CrI$_3$ due to the presence of dihedral mirrors ($\sigma_d$) and two-fold rotations about an in-plane axis ($C_2'$) that are absent in the bulk. This results in a structural $D_{3d}$ point group, which is twice the size of the structural group $S_6$ of bulk CrI$_3$. However, the newly added symmetries are all antiunitary in the presence of the FM spin ordering, so that the unitary group is again just $S_6$. 

\begin{figure}
\centering\includegraphics[width=0.9\columnwidth]{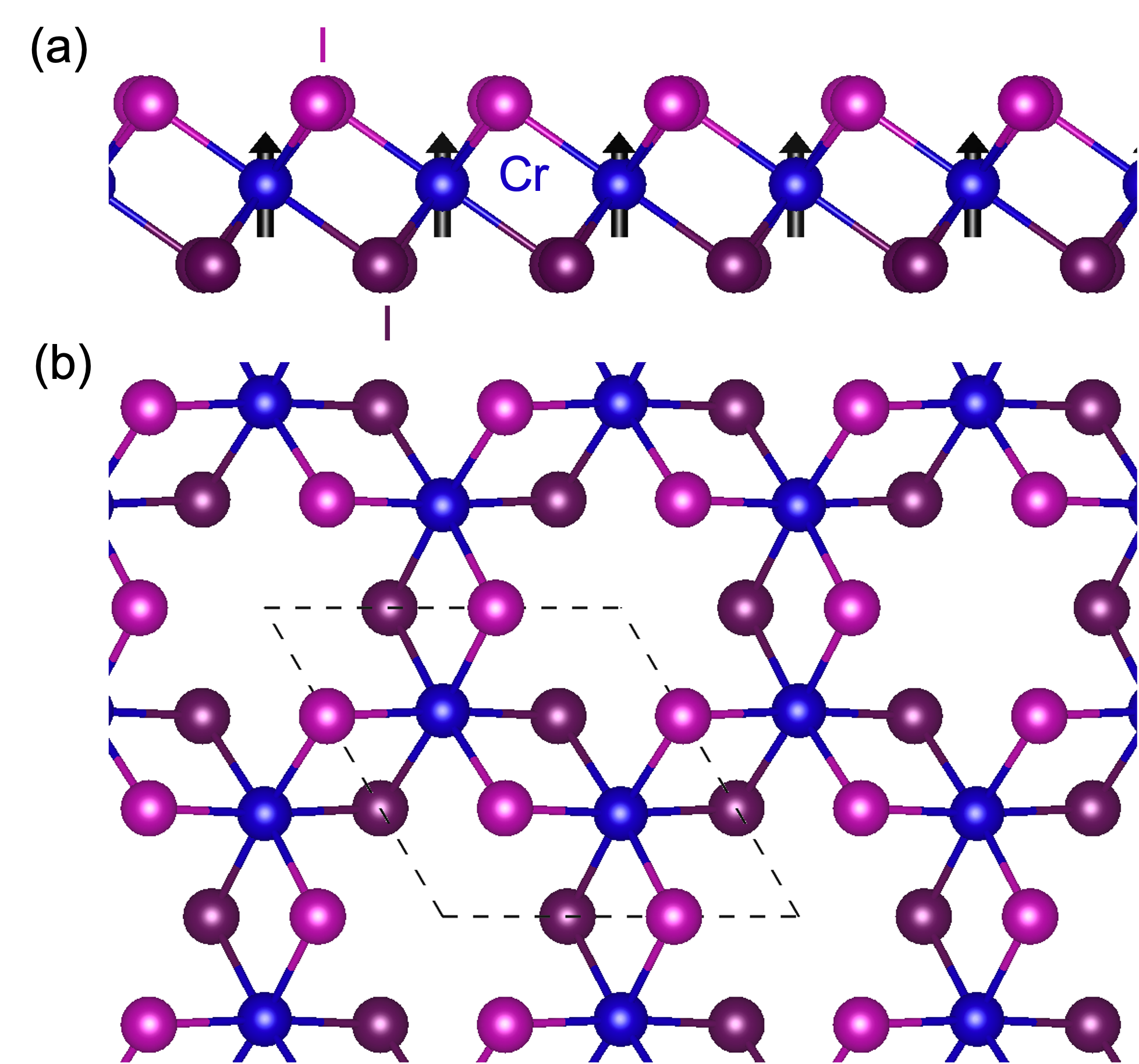}
\caption{Side (a) and top (b) views of the crystal structure of monolayer CrI$_3$, with a dashed line indicating the unit cell. Cr atoms are depicted in blue, while top-layer and bottom-layer I atoms are in bright and dark magenta, respectively. The black vectors indicate the magnetic moments, which are oriented along the $z$-direction. }
\figlab{cri3-2d-crystal}
\end{figure}

Thus, we again label the perturbed phonons at $\Gamma$ using irreps of $S_6$, under which they decompose to $4 A_g \oplus 4 E_g \oplus 4 A_u \oplus 4 E_u$, where one $A_u$ mode and two $E_u$ modes are acoustic modes. Similar to bulk CrI$_3$, the $E_g$ and $E_u$ irreps in the $S_6$ point group correspond to two 1D complex irreps rather than a single 2D irrep. As a result, the $E_g$ and $E_u$ chiral phonons are no longer doubly degenerate and have energy splittings when compared to bare phonons. We refer to the $E_g$ and $E_u$ phonons as chiral phonons because they individually respect $C_3$ symmetry, but instead of having identity $C_3$ eigenvalues, they have complex eigenvalues ($\varepsilon$ or $\varepsilon^*$). In Table~\ref{tab:res-cri3-mono-e}, we present the energies, angular momenta, and $C_3$ eigenvalues of the $E_g$ and $E_u$ chiral phonons, along with the energies of the bare phonons as a reference. Additionally, we include the results for magnons.

Comparing Tables~\ref{tab:res-cri3-mono-e} and \ref{tab:res-cri3-3d-e}, we observe that $E_g$ chiral phonons in monolayer CrI$_3$ have a similar energy splitting to those in bulk CrI$_3$, but with larger PAM because the in-plane motion has been enhanced relative to out-of-plane motion. In monolayer CrI$_3$, the $E_u$ chiral phonons around 10~meV have the largest energy splitting because they are closest in energy to the $E_u$ magnon. Overall, the mechanisms for $E_g$ and $E_u$ chiral phonons in bulk and monolayer CrI$_3$ are similar despite differences in energy splittings and magnon energies. 

% typo fixed in this table
\begin{table}
\caption{\label{tab:res-cri3-mono-e} 
Energies, $z$-direction total PAM ($L_z$), and $C_3$ eigenvalue ($\chi(C_3)$) of $E_g$ and $E_u$ chiral phonons and magnons in monolayer CrI$_3$ calculated using the full model. The bare phonon and magnon energies $E_0$ are included as a reference for the energy shift $\Delta E$. }
\begin{ruledtabular}
\begin{tabular}{lcccc}
Irrep & $E_0\mbox{~(meV)}$ & $\Delta E\mbox{~(meV)}$ & $L_z (\hbar)$ & $\chi(C_3)$ \\
\colrule
Phonons &&&& \\
\quad $E_g$ & $\phm 6.2343$ & $-0.0028$ & $\phm 0.1757$ & $\varepsilon^*$  
\\
 &  & $\phm 0.0033$ & $-0.1762$ & $\ve$  
 \\
 & $12.6725$ & $-0.0005$ & $-0.2398$ & $\varepsilon^*$  
 \\
 &  & $\phm 0.0007$ & $\phm 0.2420$ & $\ve$  
 \\
 & $13.4436$ & $-0.0010$ & $\phm 0.2234$ & $\varepsilon^*$  
 \\
 &  & $\phm 0.0011$ & $-0.2250$ & $\ve$  
 \\
 & $30.1453$ & $-0.0028$ & $\phm0.8408$ & $\varepsilon^*$  
 \\
 &  & $\phm 0.0028$ & $-0.8408$ & $\ve$   
 \\
%\colrule
\quad $E_u$ & $10.0072$ & $-0.0595$ & $\phm 0.2535$ & $\ve$ 
\\
 &  & $-0.0014$ & $-0.2314$ & $\varepsilon^*$ 
 \\
 & $14.2708$ & $-0.0063$ & $-0.7224$ & $\varepsilon^*$ 
 \\
 &  & $\phm 0.0450$ & $\phm 0.7279$ & $\ve$ 
 \\
 & $28.2112$ & $\phm 0.0010$ & $-0.9540$ & $\ve$ 
 \\
 &  & $\phm 0.0036$ & $\phm 0.9538$ & $\varepsilon^*$ 
 \\
 %\colrule
Magnons &&&& \\
 \quad $E_g$ & $\phm 0.9731$ & $-0.0069$ &  & $\ve$  
 \\
 \quad $E_u$ & $10.5988$ & $-0.0060$ &  & $\ve$ 
 \\
 \end{tabular}    
\end{ruledtabular}
\end{table}

In order to discuss the phonons belonging to the $A$ irreps, it is essential to first determine the symmetries of the bare phonons. Similar to our argument in bulk Cr$_2$O$_3$, since $\sigma_d \mathcal{T}$ and $C_2'\mathcal{T}$ remain as symmetries, the bare phonons must possess $D_{3d}$ symmetries instead of $S_6$. We have found that the $\Gamma$-point bare phonons can be decomposed into $2A_{1g} \oplus 2A_{2g} \oplus 4E_g \oplus 1A_{1u} \oplus 3A_{2u} \oplus 4E_u$, wherein one $A_{2u}$ mode and two $E_u$ modes are acoustic. The presence of magnetic moments breaks the $\sigma_d$ and $C_2'$ symmetries of the bare phonons, which leads to inter-irrep mixing between the bare phonons of different irreps in the perturbed phonons. According to the correlation table of the $D_{3d}$ point group, the $A_{1g}$ and $A_{2g}$ irreps combine to form the $A_g$ irrep of $S_6$, while the $A_{1u}$ and $A_{2u}$ irreps combine to form $A_u$. However, as in bulk Cr$_2$O$_3$, the inter-irrep mixing in monolayer CrI$_3$ is found to be relatively weak, as demonstrated by the results presented in Table~\ref{tab:res-cri3-mono-a}. This allows us to continue to refer to the perturbed phonons as, e.g., $A_{1g}$-like based on their predominant character. Nevertheless, it is important to emphasize that the correct labeling of zone-center phonons should consider the irreps of the $S_6$ group rather than the $D_{3d}$ group.

Table~\ref{tab:res-cri3-mono-a} contains the energies, angular momenta, and inter-irrep mixing of $A_g$ and $A_u$ perturbed phonons. Bare phonon energies are also included as a reference for the energy shifts. By comparing Tables~\ref{tab:res-cri3-3d-a} and \ref{tab:res-cri3-mono-a}, we observe that the perturbed phonons from the $A$ irreps in both bulk and monolayer CrI$_3$ exhibit small energy shifts and have a small PAM, indicating a weak effect of the $\gpp$ matrices in both cases. 

Concerning the magnons, we find that the energy of the optical magnon in monolayer CrI$_3$ is lower than in the bulk. This is attributable to the absence of ferromagnetic inter-layer exchange in the monolayer~\cite{sivadas2018}. In contrast, the energy for the acoustic magnon is greater than that in the bulk. As the acoustic magnon energy is related to the magnetic anisotropy energy (MAE), this suggests a greater MAE for monolayer CrI$_3$. This observation agrees with findings from another study~\cite{gudelli2019}, where the generalized-gradient approximation to the exchange-correlation functional was used~\cite{perdew1996}. The numerical values for the matrices $\kss$ and $\gss$ are provided in Appendix~\ref{ap:kss}. 

In summary, as is the case for the chiral phonons in bulk CrI$_3$, the $E_g$ and $E_u$ chiral phonons in monolayer CrI$_3$ are no longer doubly degenerate and possess significant angular momentum. The $A_g$ and $A_u$ perturbed phonons exhibit small energy shifts with respect to the bare phonons and acquire non-zero angular momentum. However, unlike the bulk case, $A_g$ and $A_u$ perturbed phonons in monolayer CrI$_3$ exhibit inter-irrep mixing due to the symmetry reduction caused by the presence of magentic order. 
From the perspective of optical activity, the $A_{2g}$-like perturbed phonons are weakly Raman-active, since they contain some components from the $A_{1g}$ sector, whereas the $A_{1u}$-like perturbed phonons acquire some IR activity due to components from $A_{2u}$.

\begin{table}
\caption{\label{tab:res-cri3-mono-a} Energies, total $z$-direction PAM ($L_z$), and inter-irrep mixing for $A$-irrep perturbed phonons in monolayer CrI$_3$. `Irrep' labels are those of the parent bare phonons, whose energies $E_0$ are included as a reference for the energy shift $\Delta E$. }
\begin{ruledtabular}
\begin{tabular}{cdddd}
Irrep & \multicolumn{1}{c}{$E_0$~(meV)} & \multicolumn{1}{c}{$\Delta E \, (10^{-8}~\mbox{meV})$} & \multicolumn{1}{c}{$L_z \,(10^{-4} \hbar)$} & \multicolumn{1}{c}{$\rho_n$ ($10^{-4}$)} \\
\colrule
$A_{1g}$ &  9.47  &  -6.3  & -2.07  & 1.47 \\
         & 16.12  & -67.4  &  1.50  & 2.30 \\
%\colrule 
$A_{2g}$ & 10.92  &   2.2  &  2.22  & 1.72 \\
         & 26.98  & 125.4  & -2.10 &  3.77 \\
\colrule
$A_{1u}$ & 16.67  &  -5.0  & -2.49  & 1.37 \\
%\colrule 
$A_{2u}$ & 7.02   &  -3.7  &  0.94  & 0.48 \\
         & 32.69  &   27.0  &  0.49  & 1.49 \\
\end{tabular}    
\end{ruledtabular}
\end{table}

%=================================================
\subsubsection{Phonons in monolayer VPSe$_3$}
\seclab{afm-2d}
%=================================================

In this section, we investigate the phonons of a VPSe$3$ monolayer, which is a 2D AFM insulator. It is predicted to have a N\'eel-type AFM structure~\cite{chittari2016}, with the magnetic moment oriented along the $z$-direction. The crystal structure of monolayer VPSe$_3$ is depicted in \fref{vpse3-crystal}. The magnetic space group is $P\bar{3}'1m$, which is again a Type-III black-white group.  The structural point group is $D_{3d}$, but the magnetic moments from V atoms break both inversion ($i$) and two-fold rotational ($C_2'$) symmetries, reducing the unitary point group to $C_{3v}$. The three-fold rotation ($C_3$) and dihedral mirror ($\sigma_d$) remain symmetries of the magnetic group, together with operations $i\mathcal{T}$, $C_2'\mathcal{T}$, and $S_6\mathcal{T}$. 

\begin{figure}
\centering\includegraphics[width=\columnwidth]{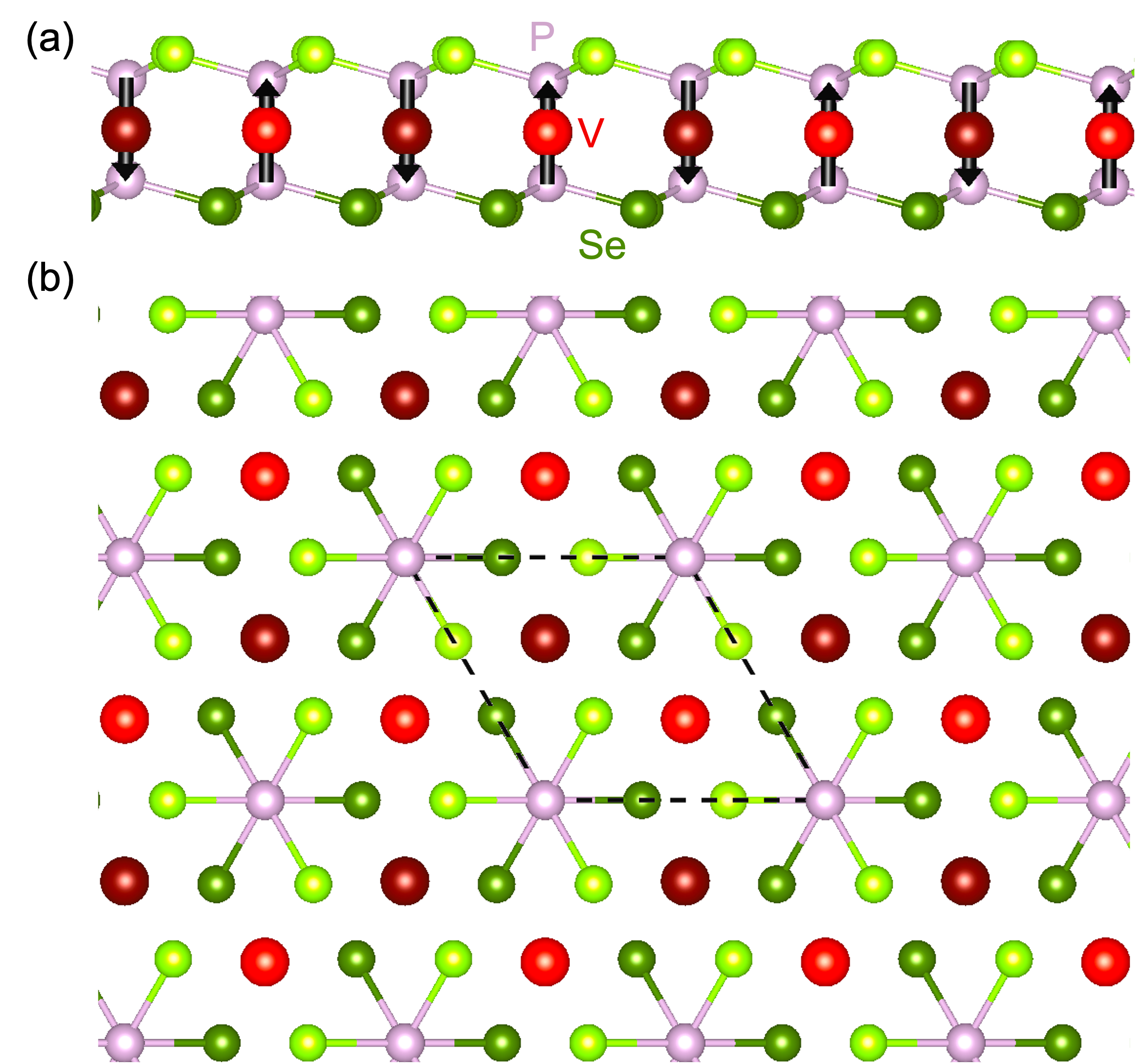}
\caption{Side (a) and top (b) views of the crystal structure of monolayer VPSe$_3$, with a dashed line indicating the unit cell. Bright and dark shades denote V atoms with spin up and down, respectively. Purple atoms represent P atoms, while top-layer and bottom-layer Se atoms are in bright and dark green, respectively. The magnetic moments, oriented along the $z$-direction, are denoted by black vectors.}
\figlab{vpse3-crystal}
\end{figure}

According to the character table for the $C_{3v}$ point group, the $E$ irrep is a true 2D irrep, i.e., not a complex conjugate pair of 1D irreps. Therefore, perturbed phonons belonging to $E$ irreps must be doubly degenerate, which is confirmed by the numerical results in Table~\ref{tab:res-vpse3}. The energies of the perturbed phonons are described by energy shifts $\Delta E$ with respect to the bare phonon energies ($E_0$).

The VPSe$_3$ monolayer maintains $i\mathcal{T}$ symmetry, similar to Cr$_2$O$_3$. This symmetry forbids the existence of a nonzero total PAM but permits the presence of nonzero ARPAMs. The magnetic space group of VPSe$_3$ is $P\bar{3}'1m$, with V, P, and Se atoms occupying the $2c$, $2e$, and $6k$ Wyckoff positions, respectively. We find that V atoms possess only nonzero $L_z$, Se atoms can only have nonzero in-plane $\bf L$, and P atoms cannot carry angular momentum. The ARPAMs for the perturbed phonons, traced over the subspace of degenerate doublet for phonons with $E$ irreps, are provided in Table~\ref{tab:res-vpse3}. It is possible to further decompose the doublets from $E$ irreps into two chiral phonons, which individually respect the $C_3$ symmetry but have different complex eigenvalues. 

Building on our earlier discussion, similar to monolayer CrI$_3$ and Cr$_2$O$_3$, the $D_{3d}$ symmetry is preserved by $\kpp$ and $\Mp$ in monolayer VPSe$_3$, allowing us to use irreps of $D_{3d}$ to label zone-center bare phonons. These bare phonons decompose to $5E_g \oplus 5E_u \oplus 3A_{1g} \oplus 4A_{2u} \oplus 1A_{1u} \oplus 2A_{2g}$, with one $A_{2u}$ and two $E_u$ modes being acoustic modes. However, due to the presence of magnetic moments on V atoms, $i$ and $C_2'$ symmetries are broken, reducing the symmetry to $C_{3v}$ and requiring the use of irreps from this point group to label perturbed phonons instead of $D_{3d}$. The inter-irrep mixing resulting from the symmetry reduction can be determined from the correlation table, which shows that $E_g$ and $E_u$ bare phonons form $E$ perturbed phonons, $A_{1g}$ and $A_{2u}$ bare phonons form $A_1$ perturbed phonons, and $A_{2g}$ and $A_{1u}$ bare phonons form $A_2$ perturbed phonons. As in the cases of monolayer CrI$_3$ and Cr$_2$O$_3$, the inter-irrep mixing caused by the magnetic order is small. Thus, we still refer to the perturbed phonons as $E_g$-like, etc., but again the correct labeling of those perturbed phonons should involve the irrep of $C_{3v}$ rather than those of $D_{3d}$. 

The inter-irrep mixing $\rho$ is defined in the same manner as with Cr$_2$O$_3$ using \eq{rho-irrep}, and the results are presented in Table~\ref{tab:res-vpse3}. The inter-irrep mixing leads to anomalous Raman/IR activities as well. For example, the $A_{1g}$-like and $E_g$-like perturbed phonons, which are strongly Raman-active, now acquire small IR activities, while IR-active $A_{2u}$-like and $E_u$-like perturbed phonons acquire small Raman activities. The $A_{2g}$-like and $A_{1u}$-like perturbed phonons remain silent in both Raman and IR activities. 

\begin{table}
\caption{\label{tab:res-vpse3} Bare phonon irreps, energies, atom-resolved phonon angular momentum (ARPAM) of V atoms in the $z$-direction (V $L_z$), ARPAM of Se atoms in the $y$-direction (Se $L_y$), and inter-irrep mixing ($\rho$) of perturbed phonons in monolayer VPSe$_3$. Bare phonon energies $E_0$ are included as a reference for energy shifts $\Delta E$. The $E$-irrep phonons are doubly degenerate. The energy shifts $\Delta E$ for $A$-irrep perturbed phonons are on the order of $10^{-5}\;\mu\mbox{eV}$.}
\begin{ruledtabular}
\begin{tabular}{cddddd}
    Irrep & \multicolumn{1}{c}{$E_0$} & \multicolumn{1}{c}{$\Delta E$} & \multicolumn{1}{c}{\textrm{V} $L_z$} & \multicolumn{1}{c}{\textrm{Se} $L_y$} & \multicolumn{1}{c}{$\rho$} \\
    & \multicolumn{1}{c}{\textrm{(meV)}} & \multicolumn{1}{c}{($\mu$\textrm{eV})} & \multicolumn{1}{c}{$(10^{-4}\hbar)$} &  \multicolumn{1}{c}{$(10^{-4}\hbar)$} & \multicolumn{1}{c}{$(10^{-4})$} \\
\colrule
$ E_g$   &  12.790 &   0.052 &  -0.238 &   0.008 &   0.384  \\
         &  15.102 &   0.230 &   0.337 &   0.707 &   4.161  \\
         &  20.127 &   0.049 &  -0.055 &  -0.441 &   1.116  \\
         &  27.728 &  -0.001 &  -9.280 &  -0.145 &   6.013  \\
         &  53.837 &   0.016 &  -0.055 &  -0.323 &  26.994  \\
%\colrule
$ E_u$   &  14.526 &   0.033 &  -0.453 &  -0.620 &   4.269  \\
         &  18.125 &   0.049 &   0.077 &   0.189 &   1.029  \\
         &  30.765 &  -0.005 &  11.104 &   0.171 &   6.768  \\
         &  53.697 &  -0.019 &   0.039 &   0.245 &  26.945  \\
\colrule
$A_{1g}$ &  18.743 &   0.000 &       0 &  -0.224 &   1.154  \\
         &  25.490 &   0.000 &       0 &   0.050 &   1.032  \\
         &  62.327 &   0.000 &       0 &  -0.012 &   0.357  \\
%\colrule     
$A_{2u}$ &  16.287 &   0.000 &       0 &   0.280 &   1.101  \\
         &  32.081 &   0.000 &       0 &  -0.104 &   0.624  \\
         &  37.856 &   0.000 &       0 &  -0.142 &   0.859  \\
\colrule     
$A_{1u}$ &  16.625 &   0.000 &       0 &       0 &   0.581  \\
%\colrule     
$A_{2g}$ &   8.955 &  -0.000 &       0 &       0 &   0.307  \\
         &  30.688 &   0.000 &       0 &       0 &   0.220  \\
\end{tabular}    
\end{ruledtabular}
\end{table}

In addition, we have studied the energies of bare and perturbed magnons. Both bare and perturbed magnons are doubly degenerate as they belong to the $E$ irrep. The bare magnon energy is found to be $8.58$~meV, while the energy shift of the perturbed magnon is $-0.81\,\mu$eV, which is more significant than in Cr$_2$O$_3$. This can be attributed to the larger SOC in VPSe$_3$. We further analyzed the symmetry of the perturbed magnon by decomposing it into $E_g$ and $E_u$ eigenmodes of $\kss$. Our findings reveal that $\rho(E_u)=0.9992$, while $\rho(E_g)=0.0405$, indicating that the magnon is almost entirely of $E_u$ symmetry, similar to the acoustic magnon in Cr$_2$O$_3$. The numerical values for the matrices $\kss$ and $\gss$ are provided in Appendix~\ref{ap:kss}. 

To summarize, we have computed the perturbed phonons of monolayer VPSe$3$. While bare phonons retain $D_{3d}$ symmetry, perturbed phonons only possess $C_{3v}$ symmetry due to the presence of magnetic moments. As in the case of Cr$_2$O$_3$, the $E$-irrep in $C_{3v}$ is a 2D irrep, preserving degeneracy despite the broken TRS. The total PAM of each perturbed phonon is zero due to $i\mathcal{T}$ symmetry, but perturbed phonons can possess nonzero ARPAM. Moreover, we have also investigated the energies of bare and perturbed magnons, finding that the energy shift of magnons is greater than in Cr$_2$O$_3$. Although the magnon is almost entirely of $E_u$ symmetry, the $E_g$-$E_u$ mixing in the magnon sector is more pronounced than in the phonons.

%=================================================
\section{Discussion} 
\seclab{diss}
%=================================================

%=================================================
\subsection{Role of spin-orbit coupling}
\seclab{soc}
%=================================================

Spin-orbit coupling (SOC) is essential to the physics described above.  We identify three matrices, namely $\ksp$, $\gsp$, and $\gpp$, all of which either violate TRS or mediate the interaction between phonons and magnons.  Here we show that in collinear systems such as those considered here, all three of these matrices vanish in the absence of SOC.  Since these matrices control the splittings of degenerate phonon modes, it follows that these splittings also vanish without SOC. This is demonstrated here using formal arguments and then confirmed via numerical calculations on bulk CrI$_3$.  

In the absence of SOC, the presence of global spin rotational symmetry implies that the exchange interactions are of pure Heisenberg form, $H = \sum_{\langle ij\rangle} J_{ij}\, \mathbf{S}_i \cdot \mathbf{S}_j$, where the $J_{ij}$ depend on the atomic coordinates.  Spin-phonon coupling in noncollinear systems is often described in terms of exchange striction, i.e., the first-order changes of $J_{ij}$ with atomic displacements (see, e.g., Ref.~\cite{toth2016}).  However, such variations of $J_{ij}$ do not induce any spin canting in an SOC-free collinear system.  This follows because the energy is stationary with respect to canting of any spin, since $\delta H = \sum_{\langle ij\rangle} J_{ij} [\mathbf{S}_i \cdot \delta \mathbf{S}_j + \delta \mathbf{S}_i \cdot \mathbf{S}_j]$, and $\mathbf{S}_i$ and $\delta \mathbf{S}_j$ are orthogonal in a collinear spin system. 

A complementary point of view comes from noting that the spinor wavefunctions are separable into real spatial wavefunctions with pure spin-up or spin-down character in an SOC-free collinear magnet.  This remains true as atoms are displaced, so that there is no induced spin canting.  This also explains why $\gpp$ vanishes.  For example, consider three structural configurations, a reference `0' and configurations with displacements $\delta q_i$ and $\delta q_j$. The Berry phase
\begin{equation}
\Phi_{ij} = -\Im \ln [
   \ip{\psi_0}{\psi_{\delta q_i}}
   \ip{\psi_{\delta q_i}}{\psi_{\delta q_j}}
   \ip{\psi_{\delta q_j}}{\psi_0} ]
\end{equation}
clearly vanishes, since all inner products are real.

The argument for the vanishing of $\gsp$ is slightly more subtle.  This time one of the displacements, say $\delta q_j$, is replaced by a spin canting $\delta s_j$ of one spin. The spin system is no longer collinear, but it is still coplanar.  In this case the spinors can be represented using Pauli matrices $\sigma_3$ and $\sigma_1$ to span the plane in which the spins lie. Both of these Pauli matrices are real, so the overall spatial-spinor wavefunctions remain real, and the Berry phase around the loop remains zero.

\begin{figure}
\centering\includegraphics[width=\columnwidth]{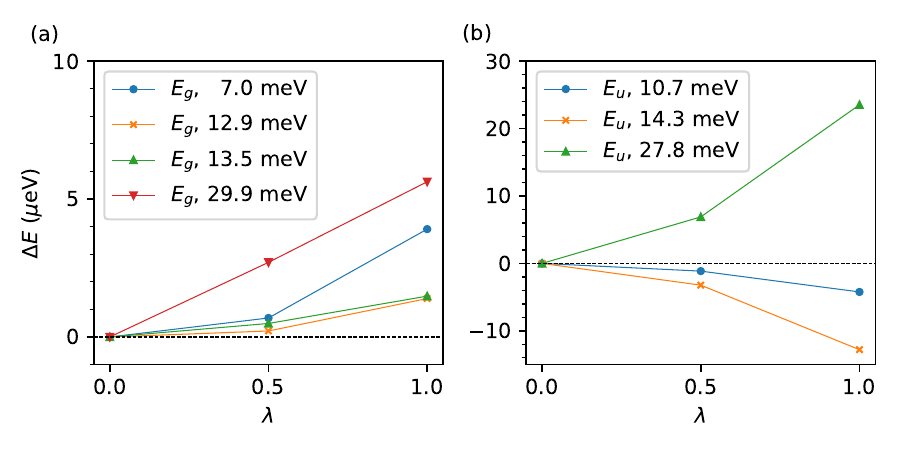}
\caption{Energy splittings ($\Delta E$) for the $E_g$ chiral phonon modes (a) and $E_u$ chiral phonon modes (b) in bulk CrI$_3$, plotted as a function of the spin-orbit coupling strength scaling factor ($\lambda$). The $\Delta E$ represents the energy difference between the `$+$' mode and the `$-$' mode near that energy.}
\figlab{e-soc}
\end{figure}

To confirm these conclusions, we repeated the calculation of the spin splittings in bulk CrI$_3$ while varying the strength of the SOC used in the first-principles calculations.  This adjustment required the introduction of a multiplicative factor in the computational code, followed by recompilation. \Fref{e-soc} shows the variation of the phonon energy splittings $\Delta E$ with the scaling factor $\lambda$ for the SOC, where $\lambda\!=\!1$ is the physical strength.  Clearly the splittings get smaller as $\lambda$ is reduced and vanish at $\lambda\!=\!0$, consistent with our expectations.

We find that $\ksp$, $\gsp$ and $\gpp$ have linear as well as quadratic terms in $\lambda$ (with the exception of the $\ksp$ coupling to acoustic magnons). The quadratic terms are dominant in most cases in \fref{e-soc}, although the linear behavior dominates for the highest-energy $E_g$ mode.

We emphasize that these conclusions hold for collinear spin systems.  While we have not studied noncollinear systems in this work, we fully expect the present approach to apply in that case.  In particular, exchange striction will induce bilinear couplings between atomic displacements and spin cantings as captured by our spin-phonon Hession $\ksp$, even in the absence of SOC. The role of SOC in the spin-phonon coupling of noncollinear systems may be an interesting avenue for future research.

%=================================================
\subsection{Connection to experiment}
\seclab{out}
%=================================================

Before concluding this paper, we would like to provide some comments on experiments. In materials where chiral phonons exhibit an energy splitting, it is possible to measure this splitting using Raman or IR spectroscopy. Since the splittings for $E_g$ chiral phonons are usually small, it may be more feasible to measure splittings of $E_u$ chiral phonons using IR spectroscopy. Additionally, circular-polarized photons with different handedness can excite chiral phonons with different chiralities, and observing the peak shift on the spectrum measured using different circular-polarized photons is possible. Another approach is to measure the magnetic moment of the excited chiral phonon using circular-polarized photons, which can be applicable regardless of the presence of energy splitting. Light-induced demagnetization has been observed in many materials, including bulk CrI$_3$~\cite{padmanabhan2022cri3}, due to electronic excitation of crystal field levels. However, by adjusting the energy of photons to resonate with a chiral phonon but not to excite electrons, one can selectively probe the chiral phonons because the energy differences between crystal field levels or semicore levels are typically larger than phonon energies. Therefore, electronic excitations can be avoided when probing chiral phonons selectively. 

%=================================================
\section{Conclusion}
\seclab{conc}
%=================================================

Based on a Lagrangian formulation, we developed a theoretical
formalism and computational methodology to determine adiabatic
dynamics in systems with multiple slow degrees of freedom. Our
computational methodology is based on static constrained DFT
calculations of Hessians and Berry curvatures with respect to the slow
parameters in order to extract the semiclassical dynamics. This
constitutes a general and computationally efficient approach that does
not require explicit time-dependent \textit{ab initio} calculations as
in, e.g., TDDFT, or other specialized capabilities beyond those
commonly found in, or easily added to, widely available DFT codes.
We demonstrate the utility of this methodology by applying it to phonons and spins
(magnons) in magnetic insulators, as the dynamics of each
is generally on the same energy scale. This represents a more
systematic way of treating spin-phonon dynamics compared to
conventional approaches which rely on building spin models and
parameterizing their dependence on atomic displacements.

Specifically, we showed that the inclusion of the Hessians and Berry
curvatures involving spin and phonon DOF are required
to accurately describe the often-neglected effects of TRS breaking on
phonon modes at the zone center. Results for four case-study materials
were presented, covering both FM and AFM ordering, as well as 2D and
3D materials. FM CrI$_3$ (both 3D bulk and 2D monolayer) exhibits
energy splittings of its $E$-type modes (which are doubly degenerate
under TRS) at the zone center. This results in chiral phonons
exhibiting circular atomic motion leading to a significant phonon
angular momentum. In AFM Cr$_2$O$_3$ (3D bulk) and VPSe$_3$ (2D
monolayer), the $E$-type modes remain doubly degenerate, but the inversion symmetry breaking from the magnetic order results in modes with mixed $E_g$ and $E_u$ character. The angular momentum of a phonon in these materials is zero in total, but can exhibit well-defined
nonzero \emph{atom-resolved} contributions.
In addition, the mixing of $E_g$ and
$E_u$ modes implies that the Raman modes acquire some small IR character and vice versa, a feature that is open to experimental confirmation. 

This work opens up various directions for future study. 
The general methodology will be useful in any material with interesting dynamics
involving multiple slow DOF. Specifically in the area
of spin-phonon dynamics, it allows the search for materials with
large splittings of chiral modes, due for example to large SOC or
approximate degeneracies between phonon and magnon branches.
It also motivates the exploration of dynamics of materials with more complicated
magnetic structures, such as noncollinear or so-called altermagnetic \cite{mazin2022,smejkal2022} orders. Finally, we focus
here on zone-center modes, but the methodology shows promise for being
generalized for computing the coupled spin-phonon dynamics across the Brillouin zone, which is relevant for the thermal Hall effect.

Overall, we expect that the developments in this work will allow
efficient and accurate calculations of generalized adiabatic dynamics,
and thus the exploration of much resulting novel physical phenomena,
in spin-phonon coupled systems and beyond.

%=================================================
\section*{Acknowledgments}
%=================================================
This work was supported by
NSF Grant DMR-1954856 (D.V.) and Grant No.~DMR-2237674 (C.E.D).
M.S. acknowledges support from Ministerio de Ciencia e Innovación (MCIN/AEI/10.13039/501100011033) through
Grant No.~PID2019-108573GB-C22; from Severo Ochoa FUNFUTURE center of excellence (CEX2019-000917-S); from Generalitat de Catalunya (Grant No.~2021 SGR 01519); and from the European Research Council (ERC) under the European Union's Horizon 2020 research and innovation program (Grant Agreement No.~724529). 
The authors thank Sinisa Coh for helpful discussions in the early stage of this work. 
S.R.~and D.V.~acknowledge the valuable discussions with Girsh Blumberg and Shangfei Wu, as well as their kind permission to reference their unpublished Raman measurement data. S.R.~acknowledges the support of Predoctoral Researcher Program of Center for Computational Quantum Physics at the Flatiron Institute. The Flatiron Institute is a division of the Simons Foundation.

\appendix
%=================================================
\section{Derivation of phonon Hamiltonian}
\aplab{ham}
%=================================================

In this section, we will review the adiabatic theory of phonons with TRS breaking, which is based on a Hamiltonian formalism first introduced by Mead and Truhlar (MT)~\cite{mead1979}. In the MT approach, the only slow variables are nuclear coordinates. One can start from the full Hamiltonian for a system with electrons and nuclei, which is~\cite{grosso2013}
\begin{align}
    & H_{\rm tot} = T_\rN(R) + T_\re(r) + V(r,R) \,, \nn
    & T_\rN(R) = - \sum_I \frac{\hbar^2 \nabla_I^2}{2 M_I} \,, \nn
    & T_\re(r) = - \sum_i \frac{\hbar^2 \nabla_i^2}{2 m} \,, 
\end{align}
where $T_\rN(R)$ and $T_\re(r)$ are nuclear kinetic energy and electronic kinetic energy, respectively. $V(r,R)$ represents the total electrostatic interactions between electrons, nuclei, and electron-nuclei interactions, which includes the effects of spin-orbit coupling. $R$ is the nuclear coordinate, and $r$ is electronic. $I$ runs over nuclei, and $i$ runs over electrons. $Z_I e$ is the positive charge for the nucleus $I$. The Schr\"odinger equation for the wave function $\Psi(r,R)$ is 
\begin{equation}
    [ T_\rN(R) + T_\re(r) + V(r,R)] \Psi(r,R) = W \Psi(r,R) \,
    \eqlab{se}
\end{equation}
where $W$ is the energy for the whole system. 

Now one can implement the Born-Oppenheimer approximation~\cite{born1954}, which treats $R$ as a slow variable and assumes the electronic dynamics is always much faster than nuclear vibrations. As a consequence, the wave function can be separated to electronic and nuclear part  
\begin{equation}
    \Psi(r,R) = \psi(r;R) \chi(R) \,,
    \eqlab{bo-wf}
\end{equation}
where the electronic part $\psi(r;R)$, which satisfies
\begin{equation}
    [T_\re(r) + V(r,R)] \psi(r;R) = \epsilon(R) \psi(r;R) \,,
    \eqlab{bo-se}
\end{equation}
is the normalized ground state wave function with respect to the nuclear coordinates $R$. $\psi(r;R)$, $V(r,R)$ and $\epsilon(R)$ depends on $R$ parametrically. One should also notice that $\psi(r;R)$ has a $U(1)$ gauge freedom, namely 
\begin{equation}
    \tilde{\psi}(r;R) = e^{i\phi(R)} \psi(r;R)
    \eqlab{psi-gauge}
\end{equation}
is also a solution to \eq{bo-se}, where $e^{i\phi(R)}$ is an $R$-dependent phase factor. Plugging \eq{bo-wf} into \eq{se}, multiplying $\psi^*(r;R)$ from the left, and integrating over $r$, gives
\begin{align}
    & \left[\sum_{I\alpha} \frac{(P_{I\alpha}-\hbar A_{I\alpha})^2}{2M_I} + \epsilon(R) + \Lambda(R)\right] \chi(R) = W\chi(R) \,, \nn
    & \Lambda(R) = \frac{\hbar^2}{2M_I} \bigl( \ip{\partial_{I\alpha} \psi(R)}{\partial_{I\alpha} \psi(R)} - A_{I\alpha}^2 \bigl) \,, \nn 
    & A_{I\alpha} = i\ip{\psi(R)}{\partial_{I\alpha} \psi(R)} \,, 
\end{align}
where $A_{I\alpha}$ is a Berry potential with respect to nuclear displacements thus called ``nuclear Berry potential''. As first pointed out by Mead and Truhlar~\cite{mead1979}, it is not always possible to make $A_{I\alpha}$ zero by tuning the $R$-dependent phase $\phi(R)$ in \eq{psi-gauge}. Here the Dirac bra-ket notation is only for the electronic degree of freedom $r$. Note that $\Lambda(R)$ can be rewritten as 
\begin{align}
    &\Lambda(R) = \frac{\hbar^2}{2M_I} \me{\partial_{I\alpha} \psi(R)}{Q}{\partial_{I\alpha} \psi(R)} \,, \nn 
    & Q = 1-\ket{\psi(R)}\bra{\psi(R)} \,,
\end{align}
therefore $\Lambda(R)$ is gauge-independent. $\Lambda(R)$ is related to the expectation value of electronic kinetic energy $\ev{T_\re(r)}$ to the order of $m/M$, where $m$ and $M$ are electronic and nuclear mass, respectively. As $\ev{T_\re(r)}$ is included in $\epsilon(R)$, we have $\Lambda(R) \ll \epsilon(R)$. Since we focus on broken time-reversal symmetry (TRS) in this work, it is worth noting that $\Lambda(R)$ does not break TRS. Therefore, we disregard $\Lambda(R)$ in this work. In addition, we only consider nuclear DOF in this section. To simplify the notation, we introduce a composite index $l$ for $I\alpha$, which is used to label nuclear DOF.

Now we are ready to write down the effective Hamiltonian for phonons, which is 
\begin{equation}
    H_{\rm eff} = \sum_l \frac{(P_l-\hbar A_l)^2}{2M_l} + \epsilon(R) \,, 
    \eqlab{H-eff}
\end{equation}
and the equation of motion (EOM) is
\begin{align}
    \dot{R}_l &= \frac{\partial H}{\partial P_l} = \frac{P_l - \hbar A_l}{M_l} \,,
    \eqlab{eom-h-r}
    \\
    \dot{P}_l &= -\frac{\partial H}{\partial R_l} = \sum_m \frac{P_m - \hbar A_m}{M_m} (\partial_l A_m) - \partial_l \epsilon (R) \,, \nn
    & = \hbar \sum_m \dot{R}_m \partial_l A_m - \partial_l \epsilon (R) \,,
    \eqlab{eom-h-p}
\end{align}
where $\partial_l$ denotes $\partial/\partial R_l$. Combining \eq{eom-h-p} and \eq{eom-h-r}, one can get
\begin{align}
    M_l \ddot{R}_l &= \dot{P}_l - \hbar \dot{A}_l \nn
    &= - \partial_l \epsilon(R) + \hbar \sum_m \dot{R}_m (\partial_l A_m - \partial_m A_l) \,,
    \eqlab{eom-ham}
\end{align}
where we have used the relation $\dot{A}_l(R) = (\partial_m A_l) \dot{R}_m$ in the second line. By introducing 
\begin{equation}
    G_{lm} = \hbar \Omega_{lm} = \hbar (\partial_l A_m-\partial_m A_l) \,,
    \eqlab{g-matrix}
\end{equation}
where the $\Omega_{lm}$ is the nuclear Berry curvature, which is gauge-invariant, \eq{eom-ham} can be written in a compact form as 
\begin{equation}
    M_l \ddot{R}_l = -\partial_l \epsilon(R) + \sum_m G_{lm}(R) \dot{R}_m \,.
    \eqlab{eom-ham-g}
\end{equation}
\Eq{eom-ham-g} is equivalent to \eq{eom-mt} for the case that $Q_i$ corresponds to a nuclear DOF. However, they are obtained using Hamiltonian and Lagrangian formalism, respectively. In this way, starting with a full quantum theory, a semiclassical theory of the dynamics has been derived. 

%=================================================
\section{From present approach to spin-phonon model}
\aplab{sp-model}
%=================================================

It has been mentioned in \sref{model-full} that the spin-phonon model in Ref.~\cite{bonini2023} only includes $\gss$ and neglects all other Berry curvature tensors, which we shall illustrate below. If we let $\gpp$, $\gps$ and $\gsp$ be all zero, then we get the EOM for the spin-phonon model as
\begin{align}
    & \Mp \ket{\ddot{u}} = - \kpp \ket{u}  - \kps \ket{s} \,, \nn 
    & \gss \ket{\dot{s}} =  \ksp \ket{u} + \kss \ket{s} \,. 
    \eqlab{eom-sp}
\end{align}
These equations can be solved exactly. However, to establish a connection with the minimal spin-phonon model proposed in Ref.~\cite{bonini2023}, we can derive Eq.~(6) in that reference from \eq{eom-sp}. 

We introduce the new phonon vector $\ket{v} = M^{1/2} \ket{u}$, so that $\Mp$ no longer appears in the EOM explicitly. Consequently, \eq{eom-sp} can be rewritten as  
\begin{align}
    \ket{\ddot{v}} &= - \tilde{K}^{\rm(pp)} \ket{v}  - \tilde{K}^{\rm (ps)} \ket{s} \,, 
    \eqlab{eom-sp-v} \\
    \gss \ket{\dot{s}} &=  \tilde{K}^{\rm(sp)} \ket{v} + \kss \ket{s} \,,
    \eqlab{eom-sp-s}
\end{align}
where 
$\tilde{K}^{\rm (pp)} = [\Mp]^{-1/2} \kpp [\Mp]^{-1/2}$,
$\tilde{K}^{\rm (ps)} = [\Mp]^{-1/2} \kps$, 
and $\tilde{K}^{\rm (sp)} = \ksp [\Mp]^{-1/2}$. 
Note that $\tilde{K}^{\rm (pp)}$ is the dynamical matrix. 

Now we only focus on bulk CrI$_3$, and include one bare phonon doublet and the magnon from the same irrep in the formalism. We introduce the unperturbed solution of \eqs{eom-sp-v}{eom-sp-s}, which we referred to as bare phonons and magnons, as 
\begin{align}
    & \tilde{K}^{\rm (pp)} \ket{v_\pm} = \omega_\rp^2 \ket{v_\pm} \,, 
    \eqlab{kpp-vpm} \\ 
    & \kss \ket{s_\pm} = -i \omega_{\rm m} \gss \ket{s_\pm} \,,  
    \eqlab{kss-spm}
\end{align}
where $\omega_\rp$ and $\omega_{\rm m}$ are energies for bare phonons and magnons, respectively. The $\pm$ subscript in $\ket{v_\pm}$ and $\ket{s_\pm}$ represents the left and right-hand circular polarized modes. Specifically, for the $E_g$ magnon, we have $\ket{s_\pm} = \frac{1}{2}(1, \mp i, 1, \mp i)^T$, where the first two components correspond to the $x$ and $y$ components of the reduced spin unit vector of the first Cr atom, and the last two components correspond to the second Cr atom. Similarly, for the $E_u$ magnon, we have $\ket{s_\pm} = \frac{1}{2}(1, \mp i, -1, \pm i)^T$.

If we adopt the assumption made in Ref.\cite{bonini2023}, which states that  
\begin{equation}
    G = -S \begin{pmatrix}
        0  &  1  & 0  & 0  \\
        -1 &  0  & 0  & 0  \\
        0  &  0  & 0  & 1  \\
        0  &  0  & -1 & 0  \\
    \end{pmatrix} \,,
    \eqlab{gss}
\end{equation}
we can verify that 
\begin{equation}
    \me{s_\pm}{\gss}{s_\pm} = \pm iS \,,
    \eqlab{gss-spm}
\end{equation}
regardless of whether $\ket{s_\pm}$ belongs to the $E_g$ or $E_u$ magnons. However, in our numerical calculations for CrI$_3$, we have observed that the effective spin $S$ for $E_g$ and $E_u$ magnons is slightly different. The numerical values for these effective spins are provided in Appendix~\ref{ap:kss}.

According to symmetry considerations, modes with different chiralities should not mix under any of the matrices in \eq{eom-sp}. Therefore, we have
\begin{align}
    & \me{s_\pm}{\ksp}{v_\pm}=\gamma \,, 
    \eqlab{ksp} \\
    & \me{s_\pm}{\ksp}{v_\mp}=0 \,,
\end{align}
where $\gamma$ represents the spin-phonon coupling strength. It is always possible to add a phase to either $\ket{s_\pm}$ or $\ket{v_\pm}$ in order to make $\gamma$ real. Therefore, we assume $\gamma$ to be real, resulting in $\me{s_\pm}{\ksp}{v_\pm} = \me{v_\pm}{\kps}{s_\pm}=\gamma$. 

We can expand $\ket{v}$ and $\ket{s}$ in the bases of $\ket{v_\pm}$ and $\ket{s_\pm}$ as 
\begin{align}
    & \ket{v} = x_\pm \ket{v_\pm} \,, 
    \eqlab{vpm} \\
    & \ket{s} = s_\pm \ket{s_\pm} \,,
    \eqlab{spm}
\end{align}
where $x_\pm$ and $s_\pm$ are coefficients representing the magnitudes of the contributions from the corresponding basis vectors. In the minimal spin-phonon model, it is assumed that the mixing between different $E_g$ doublets is negligible. However, in reality, this mixing, which is mediated by magnons, plays a crucial role in explaining why the angular momenta of the two circular polarized modes do not cancel out exactly. A more detailed analysis of this mixing based on the perturbation approach can be found in Appendix~\ref{ap:pert-ksp}. 

By multiplying $\bra{v_\pm}$ to \eq{eom-sp-v} and substituting Eqs.(\ref{eq:kpp-vpm}), (\ref{eq:vpm}), and (\ref{eq:ksp}), we obtain
\begin{equation}
    (\omega_\rp^2 - \omega^2 )x_\pm = -\gamma s_\pm \,.
\end{equation}
Similarly, by multiplying $\bra{s_\pm}$ to \eq{eom-sp-s} and substituting Eqs.(\ref{eq:kss-spm}), (\ref{eq:gss-spm}), and (\ref{eq:ksp}), we get
\begin{equation}
    (\pm \omega_s - \omega) s_\pm = \mp S^{-1} \gamma x_\pm \,.
\end{equation}
Thus, we have successfully derived Eq.~(6) in Ref.~\cite{bonini2023}. 

Although \eq{eom-sp} is derived from an adiabatic Lagrangian formalism, we aim to demonstrate that it can also emerge from the undamped Landau-Lifshitz equation~\cite{landau1935}. In its original form, the Landau-Lifshitz equation is presented as
\begin{equation}
    \frac{d\vec{M}}{dt} = - \gamma \vec{M} \times \frac{d H}{d \vec{M}} \,,
    \eqlab{ll-m}
\end{equation}
where $H$ is the Hamiltonian, ${d H}/{d \vec{M}}$ is the effective magnetic field, and $\gamma$ represents the gyromagnetic ratio, which is the ratio of the magnetic moment $\vec{M}$ to its corresponding angular momentum $\vec{S}$. As the dynamic variable in our study is the unit vector of spin, $\vec{s} = \vec{S}/S$, we can recast \eq{ll-m}) in terms of $\vec{s}$ as 
\begin{equation}
    S\frac{d\vec{s}}{dt} = - \vec{s} \times \frac{d H}{d \vec{s}} \,.
    \eqlab{ll-s}
\end{equation}
\Eq{ll-s} can be rewritten in component form as 
\begin{equation}
    S\frac{d s_\alpha}{dt} = - \epsilon_{\alpha\beta\gamma} s_\beta \frac{d H}{d s_\gamma} \,,
\end{equation}
where $\alpha$, $\beta$, and $\gamma$ span over the Cartesian coordinates. Given that the spin deviation from the $z$-direction is small, we can make a simplification on the right-hand side of this equation by setting $\beta = z$. Consequently, the equation transforms to 
\begin{equation}
    S\frac{d s_\alpha}{dt} = \epsilon_{\alpha\gamma} \frac{d H}{d s_\gamma} \,,
\end{equation}
where now, $\alpha$ and $\gamma$ are confined to $x$ and $y$ Cartesian directions. By reintroducing the matrix form and transferring the $\epsilon$ tensor to the left-hand side, we arrive at 
\begin{equation}
    - S \epsilon \frac{d\vec{s}}{dt} = \frac{d H}{d \vec{s}} \,.
\end{equation} 

Referring back to \eq{gss}, it is noteworthy that $-S\epsilon$ is equivalent to the $\gss$. Since the Hamiltonian has been expanded up to the quadratic order, the derivative $dH/d \vec{s}$ can be expressed as
\begin{equation}
    \frac{d H}{d \vec{s}} = \kss \ket{s} + \ksp \ket{u}  \,,
\end{equation}
where we have chosen to use the bra-ket notation for ease of representation. This gives rise to
\begin{equation}
    \gss \ket{\dot{s}} = \kss \ket{s} + \ksp \ket{u} \,,
\end{equation}
which aligns with the form presented in \eq{eom-sp}. Thus, we have successfully shown the equivalence of the EOM for the spin-phonon model presented in this Appendix with the undamped Landau-Lifshitz equation. 

%=================================================
\section{Spin-spin Hessians ($\kss$) and Berry curvatures ($\oss$) for all four materials}
\aplab{kss}
%=================================================

In this appendix, we present the spin-spin Hessians and Berry curvatures for all four materials investigated in our study. 

For bulk CrI$_3$, both the spin-spin Hessian ($\kss$) and Berry curvature ($\oss$) matrices have the form
\begin{equation} 
M_{4\times 4} = 
    \begin{pmatrix}
        a & b \\
        b & a 
    \end{pmatrix} \,,
    \eqlab{cri3-m4}
\end{equation}
where both $a$ and $b$ are $2 \times 2$ matrices that are restricted to be of the form 
\begin{equation}
    a = a_{\rm sym}  
    \begin{pmatrix}
        1 & 0 \\
        0 & 1 \\
    \end{pmatrix} 
    + a_{\rm asym}  
    \begin{pmatrix}
        0 & 1 \\
        -1 & 0 \\
    \end{pmatrix}
    \eqlab{cri3-decomp} \,,
\end{equation}
and similarly for $b$, as a result of the three-fold symmetry.
In Table~\ref{tab:all-ss} we provide the symmetric and antisymmetric parts of $a$ and $b$ for both $\kss$ and $\oss$ of bulk CrI$_3$. 
It is noteworthy that the eigenvalues of $\oss$ are $1.499$ and $1.566$ for the $E_g$ and $E_u$ modes, respectively. 
These values represent the effective spins of the respective magnon modes. They are are close to $3/2$ as expected from the nominal spin of the Cr$^{3+}$ ion, but they do differ slightly, especially for the optical $E_u$ magnon. Instead, the deviation is very small for the acoustic $E_g$ magnon. These deviations were not captured by the minimal spin-phonon model proposed in Ref.~\cite{bonini2023}. 

The symmetry-allowed matrix elements for monolayer CrI$_3$ exhibit a similar structure to those of bulk CrI$_3$, and they can also be decomposed following the same procedure as described in \eqs{cri3-m4}{cri3-decomp}. The numerical values of these matrix elements are provided in Table~\ref{tab:all-ss}.

For bulk Cr$_2$O$_3$, the non-zero matrix elements of $\kss$ are given by 
\begin{equation}
    \kss = \begin{pmatrix}
        a   & b   & c & d   \\
        b^T & a^T & d & c^T \\ 
        c^T & d^T & a & b^T \\
        d^T & c   & b & a^T \\
    \end{pmatrix} \,,
    \eqlab{ksscro}
\end{equation}
while the non-zero matrix elements of $\gss$ are 
\begin{equation}
    \gss = \begin{pmatrix}
        a    & b    & c  & d   \\
        -b^T & a^T  & d  & c^T \\ 
        -c^T & -d^T & a  & b^T \\
        -d^T & -c   & -b & a^T \\
    \end{pmatrix} \,,
    \eqlab{gsscro}
\end{equation}
Here, $a$, $b$, $c$, and $d$ can also be decomposed in the same manner as in \eq{cri3-decomp}. The numerical results for these matrix elements are provided in Table~\ref{tab:all-ss}. 

For monolayer VPSe$3$, the matrix elements of both $\kss$ and $\oss$ have the form
\begin{equation} 
M_{4\times 4} = 
    \begin{pmatrix}
        a & b \\
        b & a^T 
    \end{pmatrix} \,,
    \eqlab{vpse3-m4}
\end{equation}
where $a$ and $b$ are $2\times 2$ matrices. The values of these matrix elements can be found in Table~\ref{tab:all-ss}. 

From Table~\ref{tab:all-ss} we can find that, for the systems with at most two spins, the asymmetric (symmetric) parts of $\kss$ ($\gss$) all vanish. However, the symmetry is more complicated in Cr$_2$O$_3$, where some of the $b$, $c$, and $d$ components in \eqs{ksscro}{gsscro} have both symmetric and asymmetric components.

\begin{table}
\caption{\label{tab:all-ss} Matrix elements $\kss$ and $\oss$ for materials understudy (ML = monolayer). Symmetric (`sym') and antisymmetric (`asym') parts are defined in \eq{cri3-decomp}. For $\oss$ of Cr$_2$O$_3$, the values of $c_{\rm sym} = -6.211\times10^{-5}$ and $d_{\rm sym} = -2.409\times10^{-4}$ appear rounded to zero.} 
\begin{ruledtabular}
\begin{tabular}{lcdddd}
& & \multicolumn{2}{c}{$\kss$~(meV)} & \multicolumn{2}{c}{$\oss$} \\
Material & Term & \multicolumn{1}{c}{sym} & \multicolumn{1}{c}{asym} & \multicolumn{1}{c}{sym} & \multicolumn{1}{c}{asym} \\
\colrule
CrI$_3$     & a & 18.340 & 0 & 0 & -1.533 \\
            & b & -17.455 & 0 & 0 & 0.033 \\
ML CrI$_3$  & a & 8.956 & 0 & 0 & -1.526 \\
            & b & -7.497 & 0 & 0 & 0.026 \\
Cr$_2$O$_3$ & a & 72.182 & 0 & 0 & -1.454 \\
            & b & 17.360 & 0.082 & 0 & 0 \\
            & c & -27.193 & 1.275 & 0.000 & 0.023 \\
            & d & 27.479 & 0 & 0.000 & 0 \\
ML VPSe$_3$ & a & 143.092 & 0 & 0 & -1.351 \\
            & b & 142.622 & 0 & 0 & 0 \\
\end{tabular}    
\end{ruledtabular}
\end{table}

%=================================================
\section{Perturbation treatment of phonon-magnon dynamics}
\aplab{pert}
%=================================================

In the main text, we have directly solved the equation of motion (EOM) of the present approach (\eq{eom-full}). However, employing perturbation theory to solve the EOM can offer valuable physical insights into the influence of Hessian matrices and Berry curvatures on the results. In this section, we will present a perturbation treatment for $\ksp$, $\gps$, and $\gpp$ individually. Furthermore, we will provide numerical results from the perturbation treatment of $\ksp$ and compare them with the predictions of the spin-phonon model.

We begin by reformulating \eq{eom-full} in matrix form as
\begin{equation}
    (K + i\omega_i G - \omega_i^2 M ) \ket{q_i} = 0 \,,
    \eqlab{eom-mat}
\end{equation}
where the matrices $M$, $K$, and $G$ are defined in \eq{full-mat}.
At variance with the main text, in this Appendix we let $i,j$ label mixed modes, while $m,n$ and $\mu,\nu$ label bare phonon and bare magnon modes respectively.
We utilize the bare phonon and magnon states $\ket{u_n^{(0)}}$ and $\ket{s_\mu^{(0)}}$ as the basis functions for the perturbation theory, where the superscript `(0)' denotes the zeroth order in the perturbation expansion. 

The EOM for unperturbed phonons can be expressed in matrix form as
\begin{equation}
    (\kpp- \xi_n \Mp ) \ket{u_n^{(0)}} = 0
\end{equation}
where for convenience we have introduced the squared frequency $\xi_n = \omega_n^2$, and for the magnons as
\begin{equation}
    (\kss + i \omega_\mu \gss ) \ket{s_\mu^{(0)}} = 0 \,,
    \eqlab{smu-def}
\end{equation}
where $\omega_\mu$ may be positive or negative even though solutions with negative energy are not physically observable.
The fact that $\me{u_m^{(0)}}{\Mp}{u_n^{(0)}} = 0$ for $\xi_m \neq \xi_n$ and that $\me{s_\mu^{(0)}}{\gss}{s_\nu^{(0)}} = 0$ for $\omega_\mu \neq \omega_\nu$ allows us to adopt the normalization conditions
\begin{align}
    \me{u_m^{(0)}}{\Mp}{u_n^{(0)}} &= \delta_{mn} \,,
    \eqlab{p-norm} \\
    \me{s_\mu^{(0)}}{i\gss}{s_\nu^{(0)}} &= \delta_{\mu\nu} \sigma_\mu \,,
    \eqlab{s-norm}
\end{align}
where $\sigma_\mu$ takes the values $\mp1$ for $\omega_\mu > 0$ and $\omega_\mu < 0$ respectively.
The solution to \eq{eom-mat} is a composite vector that includes both the phonon and magnon sectors, living in the space spanned by basis vectors
$\ket{u_n^{(0)}} \oplus \ket{0}$ and
$\ket{0} \oplus \ket{s_\mu^{(0)}} $.

As discussed in the main text, phonons and magnons belonging to the $E$ irreducible representations can be doubly degenerate. To address this, we need to employ degenerate perturbation theory. However, we can simplify the analysis by considering the `$+$' and `$-$' sectors separately. Phonons and magnons from different sectors do not mix, allowing us to work with these bases throughout this section.

%=================================================
\subsection{Perturbation treatment of $\ksp$}
\aplab{pert-ksp}
%=================================================

We begin by examining the perturbation of $\ksp$ alone, neglecting
$\gpp$ and $\gsp$.  We replace $\ksp$ with $\lambda\ksp$, where $\lambda$ serves as our perturbation parameter, and expand \eq{eom-mat} up to second order in $\lambda$ as 
\begin{align}
    K &= K^{(0)} + \lambda K^{(1)} \,,\nn
    K^{(0)} &= 
    \begin{pmatrix}
        \kpp & 0 \\ 
        0 & \kss
    \end{pmatrix}\; \,,\nn
    K^{(1)} &= 
    \begin{pmatrix}
        0 & \kps \\
        \ksp & 0 
    \end{pmatrix} \,,\nn
    \omega_i &= \omega_i^{(0)} + \lambda \omega_i^{(1)} + \lambda^2 \omega_i^{(2)} + \ldots \,,\nn
    \xi_i &= \xi_i^{(0)} + \lambda \xi_i^{(1)} + \lambda^2 \xi_i^{(2)} + \ldots \,,\nn 
    \ket{q_i} &= \ket{q_i^{(0)}} + \lambda \ket{q_i^{(1)}} + \lambda^2 \ket{q_i^{(2)}} + \ldots \,.
\end{align}
We substitute these equations into \eq{eom-mat} and expand in orders of $\lambda$, making use of the relations
\begin{align}
    & \xi_i^{(0)} = (\omega_i^{(0)})^2 \,,\nn
    & \xi_i^{(1)} = 2\omega_i^{(0)}\omega_i^{(1)} \,,\nn 
    & \xi_i^{(2)} = (\omega_i^{(1)})^2 + 2 \omega_i^{(0)} \omega_i^{(2)} \,.
\end{align}
At zero order this yields
\begin{equation}
    (K^{(0)} + i\omega_i^{(0)} G - \xi_i^{(0)} M ) \ket{q_i^{(0)}} = 0 \,,
\end{equation}
which corresponds to the EOM for decoupled phonons and magnons.

The equation at first order in $\lambda$ is 
\begin{align}
    & (K^{(0)} + i\omega_i^{(0)} G - \xi_i^{(0)} M ) \ket{q_i^{(1)}} \nn 
    & \qquad + (K^{(1)} + i\omega_i^{(1)} G - \xi_i^{(1)} M ) \ket{q_i^{(0)}} = 0 \,. 
  \eqlab{pert-1st}
\end{align}
Multiplying \eq{pert-1st} on the left by $\bra{q_i^{(0)}}$, noting that $\bra{q_i^{(0)}} (K^{(0)} + i\omega_i^{(0)} G - \xi_i^{(0)} M ) = 0$, and also that $\me{q_i^{(0)}}{K^{(1)}}{q_i^{(0)}} = 0$ since $K^{(1)}$ is block-off-diagonal in the phonon and magnon DOF, we obtain
\begin{equation}
    \omega_i^{(1)} \Big( \me{q_i^{(0)}}{iG}{q_i^{(0)}} - 2\omega_i^{0} \me{q_i^{(0)}}{M}{q_i^{(0)}} \Big) = 0 \,,
\end{equation}
which indicates that $\omega_i^{(1)} = 0$, so that also $\xi_i^{(1)} = 0$. This is similar to the perturbation theory in quantum mechanics, where the first-order energy correction is always given by the diagonal matrix element of the interaction Hamiltonian, which is zero in our case. 

To obtain $\ket{q_i^{(1)}}$, we multiply \eq{pert-1st} by $\bra{q_j^{(0)}}$ on the left and use $\omega_i^{(1)} = \xi_i^{(1)} = 0$. This gives
\begin{equation}
    \me{q_j^{(0)}}{ K^{(0)} + i\omega_i^{(0)} G - \xi_i^{(0)} M }{q_i^{(1)}} + \me{q_j^{(0)}}{ K^{(1)} }{q_i^{(0)}} = 0 \,. 
    \eqlab{pert-1st-2}
\end{equation}
We observe that if $i$ and $j$ both label phonons or both label magnons, $K^{(1)}$ has no effect. Therefore, $\ket{q_n^{(1)}}$ for phonons has a pure magnon character, and $\ket{q_\mu^{(1)}}$ for magnons has a pure phonon character. 

%--------------------------------------
\subsubsection{Perturbation of phonons}
%--------------------------------------

We focus on the perturbation for phonons first, and we replace $\ket{q_i}$ and $\ket{q_j}$ in \eq{pert-1st-2} with $\ket{u_n}$ and $\ket{s_\mu}$, respectively. Then \eq{pert-1st-2} becomes
\begin{equation}
    \me{s_\mu^{(0)}}{ K^{(0)} + i\omega_n^{(0)} G - \xi_n^{(0)} M }{u_n^{(1)}} + \me{s_\mu^{(0)}}{\ksp}{u_n^{(0)}} = 0 \,.
    \eqlab{pert-1st-3}
\end{equation}
To obtain $\ket{u_n^{(1)}}$, we expand it in terms of the basis $\ket{s_\nu^{(0)}}$ as
\begin{equation}
    \ket{u_n^{(1)}} = \sum_\nu c_{\nu n}^{(1)} \ket{s_\nu^{(0)}} \,.
    \eqlab{un-1st-exp}
\end{equation}
Substituting \eq{un-1st-exp} into \eq{pert-1st-3}, we have
\begin{equation}
    \sum_\nu c_{\nu n}^{(1)} \me{s_\mu^{(0)}}{\kss+i\omega_n^{(0)}\gss}{s_\nu^{(0)}} + \me{s_\mu^{(0)}}{ \ksp }{u_n^{(0)}} = 0 \,.
\end{equation}
Using the relation $(\kss+i\omega_\nu^{(0)}\gss)\ket{s_\nu^{(0)}} = 0$, we find
\begin{align}
    &\sum_\nu c_{\nu n}^{(1)} \omega_{ n \nu }^{(0)} \me{s_\mu^{(0)}}{ i\gss }{s_\nu^{(0)}} + \me{s_\mu^{(0)}}{ \ksp }{u_n^{(0)}} \nn 
    & =  c_{\mu n}^{(1)} \omega_{ n \mu }^{(0)} \sigma_\mu + \me{s_\mu^{(0)}}{ \ksp }{u_n^{(0)}} = 0 \,,
    \eqlab{pert-ph}
\end{align}
where $\omega_{n\mu}^{(0)}$ denotes $\omega_n^{(0)} - \omega_\mu^{(0)}$. Using \eq{pert-ph}, we can determine $c_{\mu n}^{(1)}$ and express $\ket{u_n^{(1)}}$ as
\begin{equation}
    \ket{u_n^{(1)}} = \sum_\mu  \sigma_\mu \frac{\me {s_\mu^{(0)}} { \ksp } {u_n^{(0)}} } {\omega_\mu^{(0)} - \omega_n^{(0)}} \ket{s_\mu^{(0)}} \,.
    \eqlab{un-1st}
\end{equation}

This is the first major result of this Appendix.
It is worth noting that the first-order perturbation of the phonons only has spin character, so that the phonon character remains unchanged. This cannot explain why the angular momentum summed over a pair of $E_g$ or $E_u$ chiral phonons is not zero, as can be seen in Tables~\ref{tab:res-cri3-3d-e} and \ref{tab:res-cri3-mono-e} of the main text. To explain this effect, we need to consider the second-order perturbation. 

Since we have seen that the first-order correction $\omega_n^{(1)}$ to the phonon energy vanishes, we need to calculate the second-order correction $\omega_n^{(2)}$ in order to determine the energy splitting of the chiral phonons. By expanding \eq{eom-mat} to the second order in $\lambda$, we obtain
\begin{align}
    & (K^{(0)} + i\omega_i^{(0)} G - \xi_i^{(0)} M ) \ket{q_i^{(2)}} + K^{(1)} \ket{q_i^{(1)}} \nn 
  & \qquad + ( i\omega_i^{(2)} G - \xi_i^{(2)} M ) \ket{q_i^{(0)}} = 0 \,. 
  \eqlab{pert-2nd}
\end{align}
Multiplying \eq{pert-2nd} by $\bra{q_i^{(0)}}$ from the left, we obtain
\begin{equation}
    \me{q_i^{(0)}}{K^{(1)}}{q_i^{(1)}} + \me{q_i^{(0)}} {i\omega_i^{(2)} G - \xi_i^{(2)}M} {q_i^{(0)}} = 0 \,.
    \eqlab{pert-2nd-diag}
\end{equation}
For phonons, we replace $\ket{q_i}$ with $\ket{u_n}$ and substitute \eq{un-1st} into \eq{pert-2nd-diag}, yielding
\begin{equation}
    \xi_n^{(2)} = \sum_\mu \sigma_\mu \frac{ \abs{\me {s_\mu^{(0)}} { \ksp } {u_n^{(0)}}}^2 } {\omega_\mu^{(0)} - \omega_n^{(0)}}
    \eqlab{ph-en-xi}
\end{equation}
or equivalently
\begin{equation}
    \omega_n^{(2)} = \sum_\mu \sigma_\mu \frac{ \abs{\me {s_\mu^{(0)}} { \ksp } {u_n^{(0)}}}^2 } { 2\omega_n^{(0)} (\omega_\mu^{(0)} - \omega_n^{(0)}) } \,.
    \eqlab{ph-en-2nd}
\end{equation}
This is a second major result. 
The summation over $\mu$ in \eqs{un-1st}{ph-en-2nd} runs over all solutions of \eq{smu-def}, including those of negative energy. Thus, the unphysical negative-energy states can still contribute to the perturbation of the phonons.
It is worth noting that \eq{ph-en-2nd} explains why phonons from the `$+$' sector exhibit larger energy splitting. Since positive-energy magnons also belong to the `$+$' sector, the interactions between magnons and phonons are stronger in the `$+$' sector due to the smaller energy denominator. 

Furthermore, we can obtain $\ket{u_n^{(2)}}$ by left-multiplying \eq{pert-2nd} with $\bra{q_j^{(0)}}$. This yields
\begin{align}
    & \me {q_j^{(0)}} {K^{(0)} + i\omega_i^{(0)} G - \xi_i^{(0)} M } {q_i^{(2)}} + \me {q_j^{(0)}} {K^{(1)}} {q_i^{(1)}} \nn 
    & \qquad + \me {q_j^{(0)}} { i\omega_i^{(2)} G - \xi_i^{(2)} M } {q_i^{(0)}} = 0 \,. 
    \eqlab{pert-2nd-offd}
\end{align}
If $i$ labels a phonon and $j$ labels a magnon, both the second and third terms in \eq{pert-2nd-offd} are zero, implying that the first term is also zero. It follows that $\ket{q_i^{(2)}}$ can only possess phonon character.
Replacing $\ket{q_i}$ and $\ket{q_j}$ by $\ket{u_n}$ and $\ket{u_m}$ and substituting \eq{un-1st} into \eq{pert-2nd-offd} yields
\begin{equation}
    \ket{u_n^{(2)}} = \sum_{m\neq n} \sum_\mu \sigma_\mu \frac {\me {u_m^{(0)}} {\kps} {s_\mu^{(0)}} \me {s_\mu^{(0)}} {\ksp} {u_n^{(0)}} } { (\xi_n^{(0)} - \xi_m^{(0)}) (\omega_\mu^{(0)} - \omega_n^{(0)}) } \ket{u_m^{(0)}} \,,
    \eqlab{un-2nd}
\end{equation}
which represents the effect of the magnon-mediated phonon-phonon interaction at second order. It provides an explanation for the non-canceling angular momentum observed in chiral phonon pairs, as shown in Table~\ref{tab:res-cri3-3d-e} and \ref{tab:res-cri3-mono-e}. 

%--------------------------------------
\subsubsection{Perturbation of magnons}
%--------------------------------------

\begin{table}
\caption{\label{tab:cri3-pert} Comparison of perturbation approach and exact solution in the spin-phonon model. The perturbation approach provides second-order perturbations to phonon and magnon energies $E_n^{(2)} = \hbar \omega_n^{(2)}$ and first-order perturbations to phonon states $c_{\nu n}^{(1)}$. The exact solutions for phonon and magnon energy shifts $\Delta E = \hbar \Delta \omega$ and the spin component of phonon-like solutions $c_{\nu n}$ are included as benchmarks. The bare phonon and magnon energies $E_n^{(0)} = \hbar \omega_n^{(0)}$ are provided as reference values for energy perturbations or shifts. }
\begin{ruledtabular}
\begin{tabular}{lddddd}
Irrep & \multicolumn{1}{c}{$E_n^{(0)}$} & \multicolumn{1}{c}{$E_n^{(2)}$} & \multicolumn{1}{c}{$ \Delta E$} & \multicolumn{1}{c}{$c_{\nu n}^{(1)}$} & \multicolumn{1}{c}{$c_{\nu n}$} \\
     & \multicolumn{1}{c}{(meV)} & \multicolumn{1}{c}{($\mu$eV)} & \multicolumn{1}{c}{($\mu$eV)} & \multicolumn{1}{c}{$(10^{-3})$} & \multicolumn{1}{c}{$(10^{-3})$} \\
\colrule
Phonons &&&&& \\
\quad $E_g$ &   7.000  &    -1.323  &  -1.324  &  1.562   &  1.563    \\
      &          &     1.566  &   1.565  &  1.850   &  1.849    \\
      &  12.929  &    -0.545  &  -0.545  &  1.021   &  1.021    \\
      &          &     0.597  &   0.597  &  1.118   &  1.118    \\
      &  13.488  &    -0.244  &  -0.244  &  0.684   &  0.683    \\
      &          &     0.266  &   0.266  &  0.746   &  0.747    \\
      &  29.852  &    -0.001  &  -0.001  &  0.050   &  0.050    \\
      &          &     0.001  &   0.001  &  0.052   &  0.052    \\
%\colrule   
\quad $E_u$ &  10.769  &    -4.431  &  -4.453  &  2.809   &  2.823    \\
      &          &    -1.593  &  -1.597  &  1.010   &  1.012    \\
      &  14.329  &   -17.563  & -17.546  &  7.680   &  7.668    \\
      &          &    -4.030  &  -4.031  &  1.762   &  1.763    \\
      &  27.823  &    -3.506  &  -3.506  &  1.962   &  1.962    \\
      &          &    35.836  &  35.623  & 20.053   & 19.947   \\
%\colrule 
Magnons &&&&& \\
%\colrule
\quad $E_g$ &   0.590  &    -4.544  &  -4.543  &          &          \\
\quad $E_u$ &  22.864  &   -22.971  & -22.757  &          &           \\
\end{tabular}    
\end{ruledtabular}
\end{table}

We now turn to the corresponding perturbation treatment of the magnons. First, we replace $\ket{q_i}$ and $\ket{q_j}$ in \eq{pert-1st-2} with $\ket{s_\mu}$ and $\ket{u_n}$, respectively. Then, we multiply \eq{pert-1st-2} by $\ket{u_n^{(0)}}$ from the left, resulting in
\begin{equation}
    \me{u_n^{(0)}}{ K^{(0)} + i\omega_\mu^{(0)} G - \xi_\mu^{(0)} M }{s_\mu^{(1)}} + \me{u_n^{(0)}}{\kps}{s_\mu^{(0)}} = 0 \,. 
    \eqlab{pert-1st-mg}
\end{equation}
Next, we expand $\ket{s_\mu^{(1)}}$ in terms of the basis of   $\ket{u_m^{(0)}}$ as 
\begin{equation}
    \ket{s_\mu^{(1)}} = \sum_m d_{m\mu}^{(1)} \ket{u_m^{(0)}} \,.
\end{equation}
Substituting this expansion into \eq{pert-1st-mg}, we obtain
\begin{align}
    &\sum_m d_{m\mu}^{(1)} \me{u_n^{(0)}}{ \kpp - \xi_\mu^{(0)} M } {u_m^{(0)}} + \me{u_n^{(0)}}{\kps}{s_\mu^{(0)}} \nn
    & \qquad = - d_{n\mu}^{(1)} \xi_{\mu n}  + \me{u_n^{(0)}}{\kps}{s_\mu^{(0)}} = 0 \,,
    \eqlab{pert-1st-mg-2}
\end{align}
From \eq{pert-1st-mg-2}, we can determine the coefficients $d_{n\mu}^{(1)}$. Therefore, the first-order perturbation of the magnons is given by 
\begin{equation}
    \ket{s_\mu^{(1)}} = \sum_n \frac {\me{u_n^{(0)}} {\kps} {s_\mu^{(0)}}} {\xi_\mu^{(0)} -\xi_n^{(0)} } \ket{u_n^{(0)}} \,. 
    \eqlab{mg-1st}
\end{equation}

To get the second order energy perturbation for magnons, we replace $\ket{q_i}$ with $\ket{s_\mu}$ and substitute \eq{mg-1st} into \eq{pert-2nd-diag}. This leads to
\begin{equation}
    \omega_\mu^{(2)} = \sum_n \sigma_\mu \frac{\abs {\me {u_n^{(0)}} {\kps} {s_\mu^{(0)}} }^2 } {\xi_n^{(0)} - \xi_\mu^{(0)}} \,.
\end{equation}

The second order perturbation of the magnon states can be obtained in a similar way as was done for the phonons. We replace $\ket{q_i}$ and $\ket{q_j}$ with $\ket{s_\mu}$ and $\ket{s_\nu}$ and substitute \eq{mg-1st} into \eq{pert-2nd-offd}. This substitution yields
\begin{equation}
    \ket{s_\mu^{(2)}} = \sum_{\nu \neq \mu} \sum_n \sigma_\nu \frac {\me {s_\nu^{(0)}} {\ksp} {u_n^{(0)}} \me {u_n^{(0)}} {\kps} {s_\mu^{(0)}} } { (\omega_\mu^{(0)} - \omega_\nu^{(0)}) ( \xi_n^{(0)} - \xi_\mu^{(0)} ) } \ket{s_\nu^{(0)}} \,.
    \eqlab{sa-2nd}
\end{equation}

We are now prepared to present numerical results obtained using the perturbation approach. The neglect of $\gsp$ and $\gpp$ corresponds to the spin-phonon model presented in Appendix~\ref{ap:sp-model}, which can be solved exactly. Consequently, we can utilize the exact solution as a benchmark to evaluate the accuracy of the perturbation approach, and we choose bulk CrI$_3$ as the benchmark system. In Table~\ref{tab:cri3-pert}
we provide several quantities calculated using perturbed energies and states. These include the second-order-perturbed phonon energies, the first-order-perturbed phonon states represented by the coefficients $c_{\nu n}^{(1)}$ defined in \eq{un-1st-exp}, and the second-order-perturbed magnon energies. The corresponding exact results are also included in Table~\ref{tab:cri3-pert} for benchmarking purposes, and we find that the perturbation treatment reproduces the exact solutions to very good accuracy. We have performed calculations of the angular momentum $L_z$ using the second-order-perturbed phonon states, and the values obtained agree with the numbers presented in Table~\ref{tab:res-cri3-3d-e} up to the fourth decimal place. As a result, we do not include the specific numerical results for $L_z$ explicitly in the Table.

%=================================================
\subsection{Perturbation treatment of $\gsp$}
\aplab{pert-gsp}
%=================================================

The treatment of $\gsp$ in the perturbation framework follows a similar procedure as that of $\ksp$. In this case, we neglect $\ksp$ and $\gpp$. Since $\gsp$ is block 
%\dvm{off-diagonal? SR: Yes. Now changed. } 
off-diagonal, the first-order perturbation of the energy is also zero. By substituting $\ksp$ with $i\omega_n\gsp$ in \eq{un-1st}, we can obtain $\ket{u_n^{(1)}}$ due to $\gsp$. Similarly, perturbations $\omega_n^{(2)}$ and $\ket{u_n^{(2)}}$ due to $\gsp$ can be obtained by replacing $\ksp$ with $i\omega_n\gsp$ in \eqs{ph-en-2nd}{un-2nd} respectively. Perturbations of magnons can be obtained in the same manner. 

%=================================================
\subsection{Perturbation treatment of $\gpp$}
\aplab{pert-gpp}
%=================================================

In this section we explore the perturbation treatment of $\gpp$,
this time neglecting $\ksp$ and $\gsp$. Taking the perturbation to be $\lambda \gpp$, \eq{eom-mat} simplifies to
\begin{equation}
( \kpp + i \omega_n \gpp - \xi_n \Mp ) \ket{u_n} = 0 \,.
\eqlab{eom-ph}
\end{equation}
We expand each term as 
\begin{align}
    & \omega_n = \omega_n^{(0)} + \lambda \omega_n^{(1)} + \lambda^2 \omega_n^{(2)} + \ldots \,,\nn
    & \xi_n = \xi_n^{(0)} + \lambda \xi_n^{(1)} + \lambda^2 \xi_n^{(2)} + \ldots \,,\nn 
    & \ket{u_n} = \ket{u_n^{(0)}} + \lambda \ket{u_n^{(1)}} + \lambda^2 \ket{u_n^{(2)}} + \ldots \,, 
\end{align}
where $\kpp$ and $\Mp$ are of zeroth order in $\lambda$, and $\gpp$ is of first order in $\lambda$. 

Expanding \eq{eom-ph} to the first order in $\lambda$, we obtain
\begin{align}
    & (\kpp - \xi_n^{(0)} \Mp) \ket{u_n^{(1)}} \nn
    & \qquad + (i \omega_n^{(0)} \gpp - \xi_n^{(1)} \Mp ) \ket{u_n^{(0)}} = 0 \,.
    \eqlab{eom-ph-1st}
\end{align}
Multiplying \eq{eom-ph-1st} by $\bra{u_n^{(0)}}$ from the left, we find the first-order perturbation to the phonon energy as
\begin{align}
    & \xi_n^{(1)} = \omega_n^{(0)} \me {u_n^{(0)}} {i\gpp} {u_n^{(0)}} \,, \nn
    & \omega_n^{(1)} = \frac{1}{2} \me {u_n^{(0)}} {i\gpp} {u_n^{(0)}} \,.
    \eqlab{gpp-omega}
\end{align}
Multiplying \eq{eom-ph-1st} by $\bra{u_m^{(0)}}$ from the left, we obtain the first-order perturbation to $\ket{u_n}$ as 
\begin{equation}
    \ket{u_n^{(1)}} = \sum_{m\neq n} \frac { i\omega_n \me {u_m^{(0)}} {\gpp} {u_n^{(0)}} } { \xi_n^{(0)} - \xi_m^{(0)} } \ket{u_m^{(0)}} \,. 
\end{equation}

Since the first-order perturbations to energies and phonon states are non-zero in this case, we do not go beyond first order here, although generalizing to higher orders is straightforward.

\bibliography{pap}

\end{document}